\documentclass[12pt,pre,aps,twocolumn,showpacs,reprint]{revtex4-1}

\usepackage{helvet}

\usepackage{enumerate}

\usepackage{array}
\usepackage{graphicx}
\usepackage{amsfonts}
\usepackage{amssymb}
\usepackage{amsmath}
\usepackage{float}
\usepackage{color}
\usepackage{bm}

\newcommand*{\rvector}[1]{\boldsymbol{#1}}
\newcommand*{\rttensor}[1]{\boldsymbol{#1}}

\begin{document}

\title{Confined disordered strictly jammed binary sphere packings}

\author{D. Chen}


\affiliation{Department of Chemistry, Princeton University, Princeton, New Jersey 08544, USA}

\author{S. Torquato}

\email{torquato@electron.princeton.edu}

\affiliation{Department of Chemistry, Department of Physics, Princeton Institute for the Science and Technology of Materials, and Program in Applied and Computational Mathematics, Princeton University, Princeton, New Jersey 08544, USA}

\begin{abstract}
Disordered jammed packings under confinement have received considerably less attention than their \textit{bulk} counterparts and yet arise in a variety of practical situations. In this work, we study binary sphere packings that are confined between two parallel hard planes, and generalize the Torquato-Jiao (TJ) sequential linear programming algorithm [Phys. Rev. E {\bf 82}, 061302 (2010)] to obtain putative maximally random jammed (MRJ) packings that are exactly isostatic with high fidelity over a large range of plane separation distances $H$, small to large sphere radius ratio $\alpha$ and small sphere relative concentration $x$. We find that packing characteristics can be substantially different from their bulk analogs, which is due to what we term ``confinement frustration''. Rattlers in confined packings are generally more prevalent than those in their bulk counterparts. We observe that packing fraction, rattler fraction and degree of disorder of MRJ packings generally increase with $H$, though exceptions exist. Discontinuities in the packing characteristics as $H$ varies in the vicinity of certain values of  $H$  are due to associated discontinuous transitions between different jammed states. When the plane separation distance is on the order of two large-sphere diameters or less, the packings exhibit salient two-dimensional features; when the plane separation distance exceeds about 30 large-sphere diameters, the packings approach three-dimensional bulk packings. As the size contrast increases (as $\alpha$ decreases), the rattler fraction dramatically increases due to what we call ``size-disparity'' frustration. We find that at intermediate $\alpha$ and when $x$ is about 0.5 (50-50 mixture), the disorder of packings is maximized, as measured by an order metric $\psi$ that is based on the number density fluctuations in the direction perpendicular to the hard walls. We also apply the local volume-fraction variance $\sigma_{\tau}^2(R)$ to characterize confined packings and find that these packings possess essentially the same level of hyperuniformity as their bulk counterparts. Our findings are generally relevant to confined packings that arise in biology (e.g., structural color in birds and insects) and may have implications for the creation of high-density powders and improved battery designs.
\end{abstract}
\maketitle

\section{Introduction}
Frictionless hard-sphere packings in three-dimensional Euclidean space $\mathbb{R}^3$ has a venerable history because this idealized model captures the salient structural features of many complex systems such as liquids \cite{Fr01, Ja13a, Ja13b}, crystals \cite{Ch00}, glasses \cite{Ch00, Za83, Ja13a, Ja13b}, colloids \cite{To02}, granular media \cite{To02, De99}, heterogeneous materials \cite{To02, Ne01, Sa03a, Sa03b}, and powders \cite{Ol09, Zo14a, Zo14b}. A packing in $d$-dimensional Euclidean space $\mathbb{R}^d$ is defined as a large collection of nonoverlapping solid objects (particles). The packing fraction $\phi$ is the fraction of space $\mathbb{R}^d$ covered by the particles. During the last decade, the well-known Kepler conjecture that the densest way to pack equal-sized spheres in $\mathbb{R}^3$ is the face-centered-cubic (fcc) lattice (or its stacking variants) was finally proven \cite{Ha05}. Equal-sized hard-sphere systems in thermodynamic \textit{equilibrium} exhibit a first-order liquid-solid phase transition \cite{Ha86, Fr01} at the freezing point. Upon a very rapid compression of a hard-sphere liquid beyond the freezing point, the system falls out of equilibrium and follows a \textit{metastable} branch, whose end state is presumably the maximally random jammed (MRJ) state in the infinite-volume limit \cite{To00, To10a, Ri96, To10b, At13, Ja13a, Ja13b}.

Roughly speaking, MRJ packings \cite{To00, To10a} are mechanically stable packings with maximal disorder. Specifically, they contain a subset of strictly-jammed (i.e., mechanically stable) particles (backbone) that allow no simultaneous collective motion of the particles and non-volume-increasing strains of the system boundary, with the unjammed remainder (rattlers) imprisoned by the backbone and possess minimal order, as measured by a variety of order metrics \cite{To00, To10a, Tr00, Do05, At13, Ho13}. MRJ packings can be considered to be prototypical glasses \cite{To10a, Ji11a, Ch14a} because they are maximally disordered nonequilibrium structures and perfectly rigid with unbounded elastic moduli \cite{To03a}. Three-dimensional MRJ packings of equal-sized spheres have a packing fraction $\phi_{\mbox{\scriptsize{MRJ}}}\approx 0.639$ \cite{To00, To10b, At13, Do04}, and an isostatic backbone \cite{At13, Ho13}, implying that the backbone possesses the minimum number of particle contacts required for strict jamming. It is important to note that the MRJ state is a mathematically well-defined state that is distinguishable from the more ambiguous random close packing state \cite{To10a, At13}, especially in two dimensions \cite{At14}.

Sphere packings with a polydispersity in size have a richer phase diagram than their monodisperse counterparts
and enable a greater control of their structural characteristics, such as density and degree of order. Polydisperse sphere packings have served as structural models for a variety of solid states matter including high temperature and pressure phases of various binary intermetallic and rare-gas compounds \cite{De06, Ca09}, alloys \cite{De05}, solid propellants \cite{Ko01}, concrete \cite{Mi81}, and ceramics \cite{Ki76}. It has been recently suggested that the packing fraction at which the viscosity of hard-sphere suspensions diverges is closely related to the MRJ density \cite{Sp14}. Also, confined MRJ binary packings are relevant in the evolution of particle segregation in dense granular flow \cite{La15}. Using the Torquato-Jiao (TJ) packing algorithm \cite{To10b}, Hopkins \textit{et al.} \cite{Ho13} have shown that MRJ packings of binary spheres can achieve anomalously large packing fractions with a range of rattler concentrations. This structural tunability capability has implications for the design of novel granular low-porosity powders \cite{Ko01, Mi81, Ki76}.

\textit{Finite} disordered jammed packings have received considerably less attention than their \textit{bulk} counterparts, whether confined or not. Recent studies have investigated finite densest local sphere packings without boundaries in both two and three dimensions \cite{Ho10, Ho11}. Moreover, other studies have shown that boundaries modify local and large-scale packing arrangements, affecting the associated macroscopic properties of the materials as well \cite{Co93, Sc97, Fo06, Og12, Ya15}. Our focus in this work is confined disordered packings since they provide many open fundamental questions (as detailed subsequently below) and arise in practice. In planar fuel cell electrode materials, the particle packings are confined in one direction and in the vicinity of MRJ states \cite{Ab14}. Spatial constraints also arise in many packing problems in biology, including packings of organelles within cells \cite{Mi92, El01}, packings of living cells that comprise a variety of tissues \cite{To02, Ge08, Ji14, Po14}, natural photonic structures consisting of dense disordered arrangements of chitin nanoparticles \cite{Pr04}, and the spatial distribution of cancer cells among healthy cells \cite{To11, Ji11b, Ji13, Ch14b}.

\textit{Quasi} two-dimensional hard-sphere packings \cite{La04} are packings that are confined in one direction with a length scale on the order of a few to tens of particle sizes. Much of the previous work on confinement focused on the study of equal-sized hard spheres in equilibrium trapped between two parallel hard planes with plane separation distances up to five sphere diameters \cite{Sc97, Fo06, Og12}. These equilibrium studies have shed light on many fundamental questions such as freezing, glass formation, and the transition of systems between two and three dimensions \cite{Co93, Sc97, Fo06, Og12}. Various exotic phases not observed in their bulk counterparts were shown to arise, including buckled monolayer, rhombic bilayer, adaptive prism phases, etc.

Desmond and Weeks \cite{De09} have investigated confined ``random close'' packings of a 50-50 binary mixture of spheres with a small to large sphere radius ratio of 5/7 in both two and three dimensions. Their findings qualitatively demonstrate that the presence of confining walls significantly alters packing characteristics, including substantially lowering packing fractions and inducing layered structures in the vicinity of the hard walls. However, as the authors pointed out themselves, it is not clear if their algorithm produced mathematically well-defined MRJ states that are mechanically stable; more specifically, the packings generated by their algorithm could be at most collectively jammed, i.e., no collective motion of any subset of particles exists that can lead to unjamming of the packing in a non-deforming container \cite{To01, To03a}.

In this work, we focus on the generation and analysis of high-fidelity {\it isostatic} MRJ binary hard-sphere packings that are confined between two parallel hard planes separated by a distance $H$ \footnote{We refer to these generated packings as MRJ due to the ability of the TJ algorithm and the use of RSA initial conditions to produce them with high fidelity in the bulk \cite{At13, Ho13}. A detailed study verifying that these are indeed MRJ states by characterizing their structural disorder via a variety of order metrics is deferred to a future work due to the challenging task of enumerating jammed states for each combination of $H$, $\alpha$ and $x$.}. We consider \textit{binary} packings since almost all real systems possess some degree of particle size disparity. The packing characteristics depend on $H$, small to large sphere radius ratio $\alpha$, and small sphere relative concentration $x$ \cite{Ho12, Ho13}. In the bulk case, the TJ sequential linear programming algorithm \cite{To10b} has been shown to produce high-fidelity MRJ packings that are strictly jammed and isostatic \cite{To10b, At13}. Here we generalize the TJ algorithm \cite{To10b} to obtain for the first time putative MRJ packings that are exactly isostatic over a large range of plane separation distances $H$, sphere size ratios, and compositions. Note that isostatic packings confined between two parallel hard planes possess $3N_B+1$ particle-particle and particle-plane contact pairs, which we will discuss in more details in Sec. II. Some open fundamental questions that we wish to address in the present study are the following:

(1) How do confined MRJ binary sphere packings differ structurally from their bulk counterparts? For example, are they isostatic as in the bulk case \cite{Ho13} or are they hyperstatic? Are rattlers more prevalent when MRJ packings are confined relative to their bulk counterparts?

(2) What are good order metrics and structural descriptors to identify and characterize true MRJ states for confined binary sphere packings?

(3) How do packing fractions, rattler fractions and order metrics vary as functions of $H$ for binary sphere packings at different $\alpha$ and $x$?

(4) How do confined MRJ packings transition between two and three dimensions?

Our findings shed light on aspects of these open questions. Specifically, we find that the rattler fractions of confined packings are generally higher than those of their bulk counterparts. We introduce an order metric $\psi$ (defined in Sec. III), which is based on the number density fluctuations in the direction perpendicular to the hard walls, to quantify the order of packings. We observe that packing fraction, rattler fraction and disorder of MRJ packings generally increase with $H$, though exceptions exist. We also observe that when $H$ is on the order of two large-sphere diameters or less, the packings exhibit salient two-dimensional (2D) features; when $H$ exceeds about 30 large-sphere diameters, the packings approach three-dimensional (3D) bulk packings. We find that at intermediate $\alpha$ and $x$, the disorder of packings is maximized for a given $H$, as measured by the aforementioned order metric $\psi$. In addition, the confined systems tend to have more backbone spheres with fewer contacts relative to their bulk counterparts, which decrease with $H$ and exhibit smaller density fluctuations compared to corresponding Poisson point processes. We also determine to what extent the confined packings retain the large-scale hyperuniformity property of their bulk counterparts. A hyperuniform many-particle system is one that is characterized by an anomalously large suppression of long-wavelength density fluctuations \cite{To03b}. Our findings in general are relevant to confined packings that arise in biology (e.g., structural color in birds \cite{Du09} and insects \cite{Pr04}) and may have implications for the creation of high-density powders \cite{Ko01, Mi81, Ki76} and improved battery designs \cite{Ab14}.

The rest of the paper is organized as follows: In Sec. II, we discuss the generalized TJ algorithm employed to generate confined packings and the relation to determine isostaticity for these packings. In Sec. III, we employ various statistics and structural descriptors to characterize confined MRJ binary hard-sphere packings. In Sec. IV, we offer concluding remarks, and propose directions in which our work might be related and extended.

\section{Simulation Procedure}
\begin{figure}[h]
\begin{center}
$\begin{array}{c}\\
\includegraphics[width=0.35\textwidth]{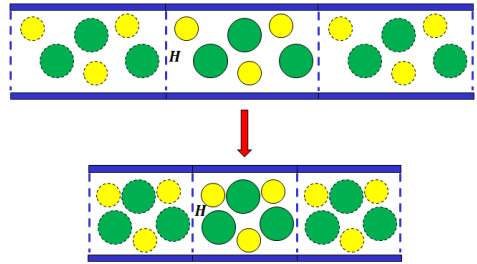}
\end{array}$
\end{center}
\caption{(Color online) Schematic illustration of the generalized TJ algorithm to generate confined MRJ binary hard-sphere packings. Periodic boundary conditions are applied to a fundamental cell in directions parallel to the fixed hard walls separated by $H$, allowing simulation box to deform and shrink on average in these directions.} \label{fig_1}
\end{figure}

In this section we first discuss the generalized TJ sequential linear programming algorithm \cite{To10b} that we will employ to generate confined packings. In our previous work \cite{To09a, To09b, To10b, At13, Ho13}, the task of generating dense packings of hard particles was formulated as an optimization problem called the adaptive shrinking cell (ASC) scheme. The objective function is taken to be the negative of the packing fraction $\phi$. The positions and orientations of the particles of fixed sizes as well as the deformation and compression/expansion of the periodic simulation box are the optimization design variables. The ASC optimization problem can be solved using various techniques, depending on the shapes of the particles \cite{To09a, To09b, To10b, At13}. In the case of sphere packings in the vicinity of a jamming basin \cite{Do04}, the objective function and impenetrability constraints can be exactly linearized. This enables one to exploit the efficient TJ sequential linear programming algorithm that enables one, in principle, to generate strictly jammed bulk sphere packings that are ordered as well as disordered, including such isostatic MRJ packings \cite{To10b, At13, Ho13, Ma13}. Here we generalize the TJ algorithm to take into account confinement and apply to produce MRJ packings of binary spheres at different $H$, $\alpha$, and $x$. It is noteworthy that although in this work the generalized TJ algorithm is employed to generate disordered jammed packings, it can be readily applied to produce confined hyperstatic ordered packings by tuning key parameters, such as the use of small number of particles and increasing the influence sphere radius \cite{To10b, Ho12}. Periodic boundary conditions are applied to a fundamental cell containing $N$ spheres in directions parallel to the fixed hard walls, allowing the simulation box to deform and shrink on average in these directions, as shown schematically in Fig. \ref{fig_1}.

The number of independent components for the strain tensor $\rttensor{\epsilon}$ applied to the fundamental cell is reduced due to the confinement and $\rttensor{\epsilon}$ possesses the following form:
\begin{equation}
\label{eq_1} \rttensor{\epsilon} = \begin{bmatrix} \epsilon_{11} & \epsilon_{12} & 0 \\ \epsilon_{12} & \epsilon_{22} & 0 \\ 0 & 0 & 0\end{bmatrix},
\end{equation}
Also, in the direction perpendicular to the planes, additional constraints are applied such that the spheres do not move beyond the impenetrable planes. Specifically, the following objective function is employed for the linear program:
\begin{equation}
\label{eq_2} \textnormal{min Tr}(\rttensor{\epsilon}) = \epsilon_{11} + \epsilon_{22},
\end{equation}
where $\rttensor{\epsilon}$ is a strain tensor that deforms and shrinks on average the fundamental cell in directions parallel to the confining planes. The adaptive fundamental cell is described by a generating matrix $\rttensor{\Lambda}$:
\begin{equation}
\label{eq_3} \rttensor{\Lambda} = \begin{bmatrix} \lambda_{11} & \lambda_{21} & 0 \\ \lambda_{12} & \lambda_{22} & 0 \\ 0 & 0 & \lambda_{33} \end{bmatrix}.
\end{equation}
The quadratic nonoverlap constraints between spheres are linearized locally to give:
\begin{equation}
\label{eq_4}
\begin{array}{c}
\rttensor{\Lambda}\cdot\rvector{r^{\lambda}_{ji}}\cdot\rttensor{\epsilon}\cdot\rttensor{\Lambda}
\cdot\rvector{r^{\lambda}_{ji}}+\Delta\rvector{x^{\lambda}_i}\cdot\rttensor{G}\cdot\rvector{r^{\lambda}_{ji}} \\
+\Delta\rvector{x^{\lambda}_j}\cdot\rttensor{G}\cdot\rvector{r^{\lambda}_{ij}} \\
\geq \frac{1}{2}[(\sigma_i+\sigma_j)/2-\rvector{r^{\lambda}_{ji}}\cdot\rttensor{G}\cdot\rvector{r^{\lambda}_{ji}}],
\end{array}
\end{equation}
where $\sigma_i$, $\rvector{x^{\lambda}_i}$ and $\Delta\rvector{x^{\lambda}_i}$ are the diameter, local coordinate (in the lattice coordinate system) and local displacement of sphere $i$, $\rvector{r^{\lambda}_{ij}}=(\rvector{x^{\lambda}_j}$+$\Delta\rvector{x^{\lambda}_j})
-(\rvector{x^{\lambda}_i}$+$\Delta\rvector{x^{\lambda}_i})$ is the relative displacement from sphere $i$ to sphere $j$, and $\rttensor{G}=\rttensor{\Lambda}^T\cdot\rttensor{\Lambda}$ is the Gram matrix of the lattice $\rttensor{\Lambda}$. The constraints that spheres cannot move beyond the fixed hard walls can be expressed as:
\begin{equation}
\label{eq_5} \frac{\sigma_i}{2\lambda_{33}} \leq (x^{\lambda}_i)_3+(\Delta x^{\lambda}_i)_3\leq1.0-\frac{\sigma_i}{2\lambda_{33}}.
\end{equation}

\subsection{Isostatic Conditions for Confinement}
We establish the relation to determine isostaticity for packing of hard spheres confined between two parallel hard planes separated by a distance $H$. For confined packings of frictionless spheres in $\mathbb{R}^3$, there are $3(N_B - 1) + 1 = 3N_B - 2$ degrees of freedom associated with translating the spheres (up to uniform translations of the whole packing under periodic boundary conditions in parallel directions), where $N_B$ is the number of (jammed) backbone spheres. The simulation box is allowed to deform for strict jamming in parallel directions and thus there are three additional degrees of freedom associated with straining the fundamental cell, totaling $F_s = 3N_B - 2 + 3 = 3N_B + 1$ degrees of freedom that must be constrained. Since the nonoverlap constraints are inequality constraints, $F_s + 1$ of them are required to satisfy isostaticity. Since the system volume cannot increase, the first constraint is $\textnormal{Tr}(\rttensor{\epsilon})\leq 0$. Therefore, the number of other constraints provided by particle-particle and particle-plane contact pairs should be equal to the number of degrees of freedom, i.e., the number of particle-particle and particle-plane contact pairs $M$ should be
\begin{equation}
\label{eq_6} M = 3N_B+1.
\end{equation}
Note that this number is different from the bulk case, where isostatic strictly-jammed sphere packings should possess $3N_B + 3$ particle-particle contact pairs \cite{At13}.

\section{Results}
We first provide some general remarks about jammed particle states in the confined space between two parallel planes separated by $H$. Not surprisingly, confined jammed packings are generally structurally distinctly different from their bulk counterparts, especially when $H$ is not much larger than the characteristic particle size $\sigma$, which is defined as the average sphere diameter. This is because certain local particle configurations found in bulk packings are inconsistent with local configurations near hard walls due to the impenetrability conditions imposed by walls. We will henceforth refer to this phenomenon as ``confinement frustration''.

\begin{figure}[h]
\begin{center}
$\begin{array}{c}\\
\includegraphics[width=0.48\textwidth]{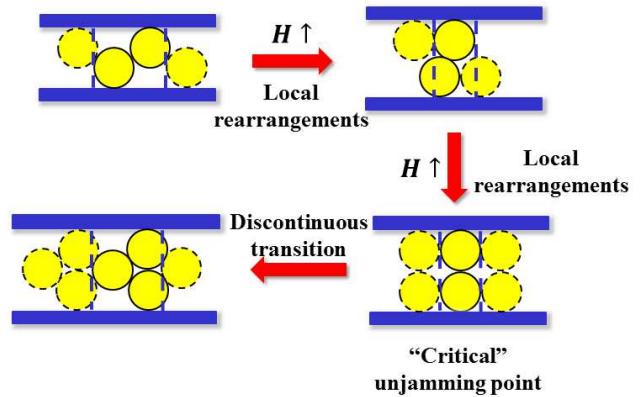}
\end{array}$
\end{center}
\caption{(Color online) Schematic illustration of the structural changes of the jammed states in the confined space between two parallel hard planes separated by $H$ as $H$ varies. For certain ranges of $H$, jammed packings remain jammed through small local rearrangements of the particles as $H$ varies, as shown in the upper panel. However, in the vicinity of certain values of $H$, a jammed packing can undergo a discontinuous transition to reach a dramatically different jammed state upon small changes in $H$, as shown in the lower panel.} \label{fig_2}
\end{figure}

Moreover, confined packings present additional subtleties because the nature of the jammed states depend on $H$ in a nontrivial way. For example, a jammed state for a fixed value of $H$ does not necessarily remain jammed upon an infinitesimal change in $H$ via infinitesimal local particle rearrangements; the packing can undergo discontinuous transitions between jammed states as $H$ varies across these critical values of $H$, which involve dramatic, finite global rearrangements of the particles, as schematically shown in Fig. \ref{fig_2}. These transitions lead to discontinuities in packing characteristics of the jammed states, as we will see later.

\begin{figure}[h]
\begin{center}
$\begin{array}{c}\\
\includegraphics[width=0.40\textwidth]{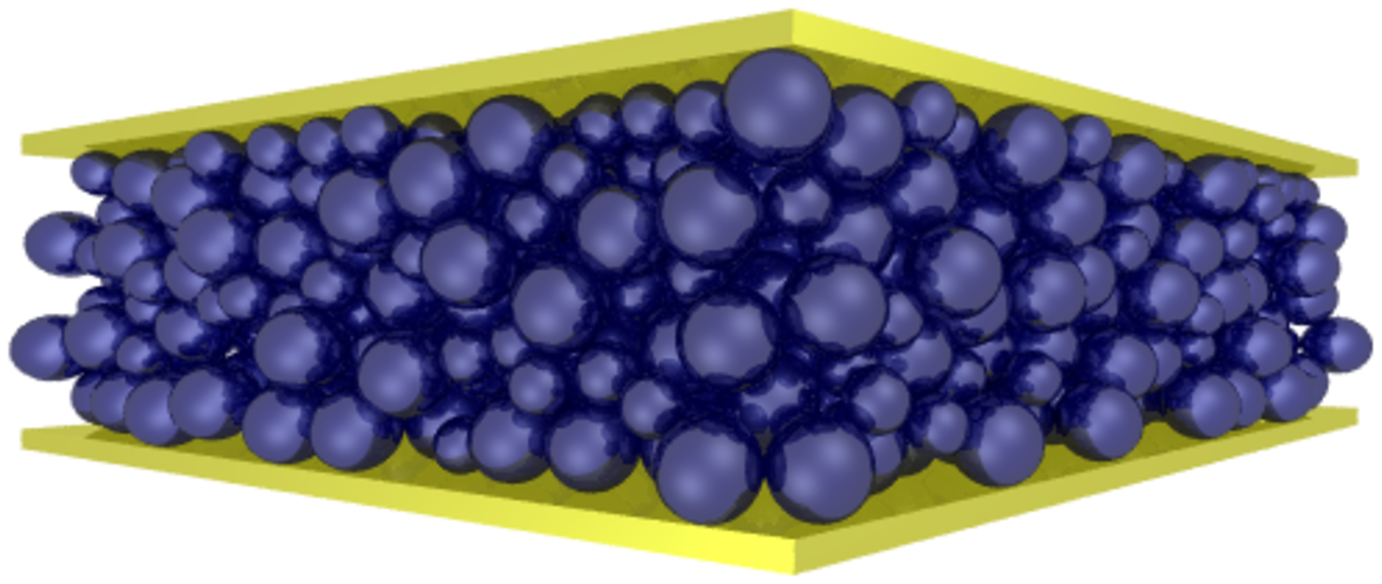} \\
\mbox{\bf (a)} \\
\includegraphics[width=0.40\textwidth]{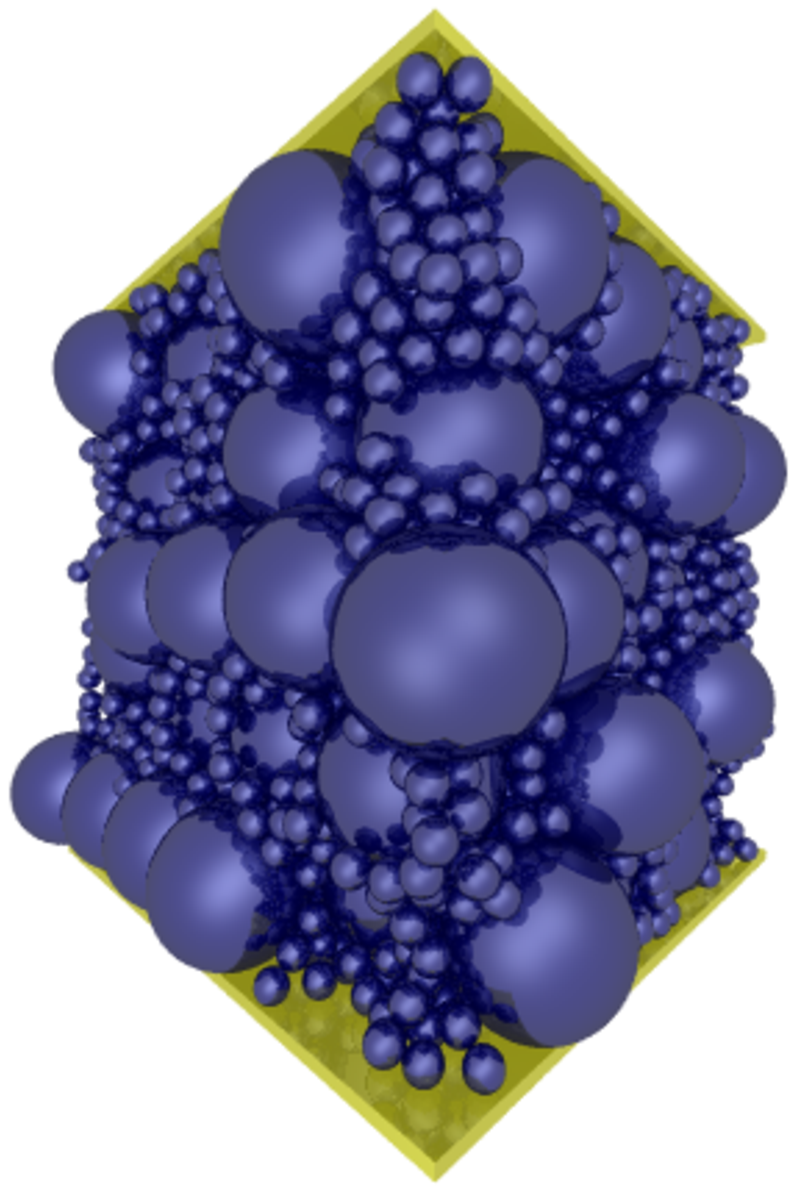} \\
\mbox{\bf (b)} \\
\end{array}$
\end{center}
\caption{(Color online) Boundaries modify local and large-scale packing arrangements, for example, inducing layered structures in their vicinity and leading to packing inefficiency. Here we present two such examples. (a) Representative MRJ binary packing of 1000 hard spheres at $\alpha = 2/3$ and $x = 0.5$ confined between two parallel hard planes at $H/\sigma = 5.0$. (b) Representative MRJ binary packing of 4000 hard spheres at $\alpha = 0.2$ and $x = 0.97$ confined between two parallel hard planes at $H/\sigma = 20.0$.} \label{fig_3}
\end{figure}

For each combination of $H$, $\alpha$, and $x$, at least 10 MRJ binary sphere packings are obtained from random sequential addition (RSA) initial conditions \cite{To02} at low initial packing fractions $0.1\leq \phi_{init} \leq 0.3$. The number of spheres $N$ in the fundamental cell is chosen such that the length scale of the generated MRJ packing in directions parallel with the hard walls is at least of the order of 10 large sphere diameters. This criterion is used to suppress finite-size effects in the directions where periodic boundary conditions are applied \cite{De09}. Specifically, at small $H$, $N$ is chosen to be 1000; while at large $H$, $N$ is chosen to be 4000. The great majority of packings that we produce using the generalized TJ algorithm are exactly isostatic according to formula (\ref{eq_6}), although a small percentage at certain values of $H$, $\alpha$, and $x$ include one or two more particle-particle and particle-plane contacts than the number corresponding to isostaticity, presumably because the numerical tolerance of the simulations were not sufficient to distinguish between proximity and contact. Figure \ref{fig_3} shows two representative packings obtained in our simulations: a 1000-sphere packing at $H/\sigma=5.0$, $\alpha=2/3$, $x=0.5$ close to the equal-sized sphere case and a 4000-sphere packing at $H/\sigma=20.0$, $\alpha=0.2$, $x=0.97$ with a large size contrast. It can be clearly seen that both configurations are disordered and densely packed, distinct from those confined crystalline phases reported previously \cite{Sc97, Fo06, Og12}. Moreover, in both cases the hard walls induce contacting layered structures that locally pack inefficiently.

\begin{figure}[h]
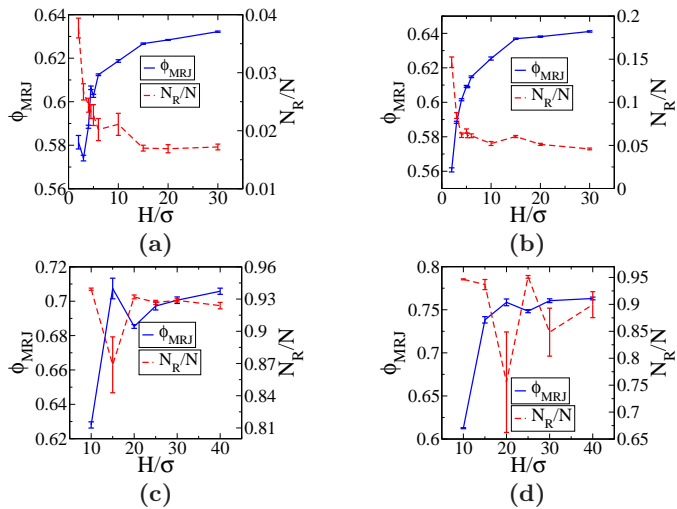

\begin{center}
$\begin{array}{c@{\hspace{1.0cm}}c}\\
\includegraphics[width=0.22\textwidth]{fig4a.eps} &
\includegraphics[width=0.22\textwidth]{fig4b.eps} \\
\mbox{\bf (a)} & \mbox{\bf (b)} \\
\includegraphics[width=0.22\textwidth]{fig4c.eps} &
\includegraphics[width=0.22\textwidth]{fig4d.eps} \\
\mbox{\bf (c)} & \mbox{\bf (d)}
\end{array}$
\end{center}
\caption{(Color online) Plot of the average packing fraction $\phi_{\mbox{\scriptsize{MRJ}}}$ and rattler fraction $N_R/N$ as functions of $H/\sigma$ obtained at various $\alpha$ and $x$, where $\sigma$ is the average sphere diameter. Each data point in the plot is obtained by averaging over 10 packings (vertical bars represent one standard deviation). (a) $\alpha = 1.0$ (equal-sized spheres). (b) $\alpha = 2/3, x = 0.5$. (c) $\alpha = 0.2, x = 0.95$. There is a sudden drop of $N_R/N$ at $H/\sigma=15.0$, which is due to a discontinuous transition between two jammed states upon small changes in $H$ in the vicinity of this dimensionless height. (d) $\alpha = 0.2, x = 0.97$. There is a sudden drop of $N_R/N$ at $H/\sigma=20.0$, which is due to a discontinuous transition between two jammed states upon small changes in $H$ in the vicinity of this dimensionless height.} \label{fig_4}
\end{figure}

For packings at different $H$, $\alpha$, and $x$, we compute their averaged packing fractions $\phi_{\mbox{\scriptsize{MRJ}}}(H; \alpha, x)$ and rattler fractions $N_R/N(H; \alpha, x)$, where $N_R$ is the number of rattlers in a packing with $N$ spheres. In Fig. \ref{fig_4} we plot $\phi_{\mbox{\scriptsize{MRJ}}}$ and $N_R/N$ as functions of $H$ at $(\alpha, x)= (1.0, -), (2/3, 0.5), (0.2, 0.95), (0.2, 0.97)$, respectively. We find that generally $\phi_{\mbox{\scriptsize{MRJ}}}$ increases with $H$ in terms of the overall trend and approaches the bulk value when $H$ is of the order of 30 large-sphere diameters, as finite-size and boundary effects become negligible. Note that the bulk value of $\phi_{\mbox{\scriptsize{MRJ}}}$ is 0.639 \cite{To10b, At13} for the monodisperse case and 0.785 \cite{Ho13} for $\alpha = 0.2$ and $x = 0.97$. We find that rattlers in confined packings are generally more prevalent than in their bulk counterparts \cite{At13, Ho13, To10b, Do05, Lu91}, which is induced by hard boundaries. As $H$ increases, the rattler fraction gradually decreases to the bulk value, e.g., roughly $1.5\%$ in the monodisperse MRJ case \cite{At13} and $14.4\%$ in the binary case of $\alpha=0.97$, $x=0.20$ \cite{Ho13}.

It is noteworthy that at certain combinations of $\alpha$ and $x$, there are local maxima and minima in $\phi_{\mbox{\scriptsize{MRJ}}}$ and $N_R/N$ as $H$ varies, for example, the local maxima for $\phi_{\mbox{\scriptsize{MRJ}}}$ and minima for $N_R/N$ at $H/\sigma=15.0, \alpha=0.95, x=0.20$ and $H/\sigma=20.0, \alpha=0.97, x=0.20$. This is due to the aforementioned discontinuous transitions. Note also that the fluctuations of $\phi_{\mbox{\scriptsize{MRJ}}}$ are less significant at $x = 0.97$ compared to $x = 0.95$. This is because when the size contrast is large, $H/\sigma_S$ is much larger than $H/\sigma_L$ ($\sigma_S$ and $\sigma_L$ are the diameters of the small and large spheres, respectively), and the hard walls affect the large spheres substantially more than the small spheres. As a result, as the number of small spheres increases, i.e., $x$ increases, the effects of the hard walls on the packing structures become less significant, and $\phi_{\mbox{\scriptsize{MRJ}}}$ depends relatively less sensitively on $H$.

In addition, as the size contrast increases, the rattler fraction dramatically increases. This is due to ``size-disparity'' frustration; i.e., in this binary system, it is not possible for a subset of spheres with the same size, surrounded by spheres with another size, to be arranged into an inherent structure (mechanically stable configurations at the local maxima in the density landscape) of the corresponding single-component system. This size-disparity frustration induces more rattlers relative to the monodisperse case, e.g., an increase as high as about $91\%$ in the rattler fraction at $H/\sigma = 15.0, \alpha = 0.2, x = 0.97$ compared to the corresponding monodisperse case. To sum up, one can see that $\phi_{\mbox{\scriptsize{MRJ}}}$ and $N_R$ depend sensitively on $H$, $\alpha$ and $x$ for $H$ comparable to sphere sizes.

\begin{figure}[h]
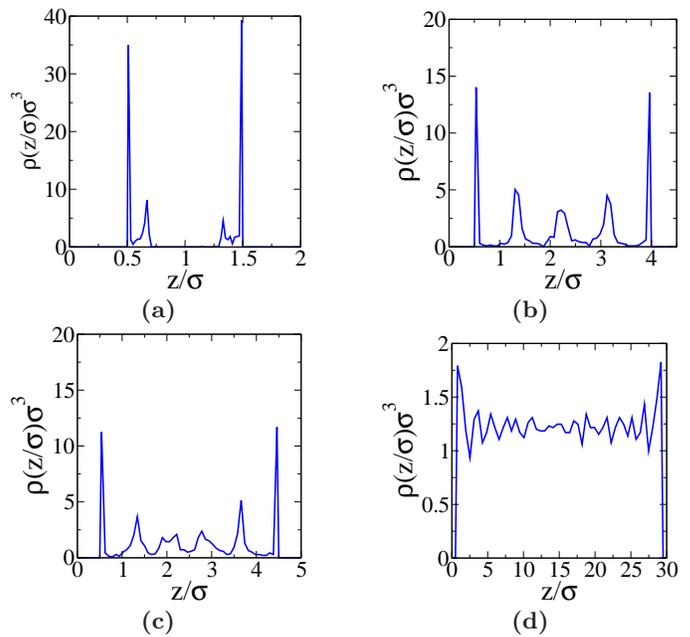

\begin{center}
$\begin{array}{c@{\hspace{1.0cm}}c}\\
\includegraphics[width=0.22\textwidth]{fig5a.eps} &
\includegraphics[width=0.22\textwidth]{fig5b.eps} \\
\mbox{\bf (a)} & \mbox{\bf (b)} \\
\includegraphics[width=0.22\textwidth]{fig5c.eps} &
\includegraphics[width=0.22\textwidth]{fig5d.eps} \\
\mbox{\bf (c)} & \mbox{\bf (d)}
\end{array}$
\end{center}
\caption{(Color online) Density profiles of confined equal-sized hard-sphere packings at various $H/\sigma$, where $\sigma$ is the average sphere diameter. (a) $H/\sigma = 2.0$. (b) $H/\sigma = 4.5$. (c) $H/\sigma = 5.0$. (d) $H/\sigma = 30.0$.} \label{fig_5}
\end{figure}

In order to further characterize the packings, we compute the number density profiles $\rho(z)$ \cite{Co93, De09} as a function of the height $z$ for different values of $H$, $\alpha$, and $x$. Specifically, we divide the space available to the sphere centers into $N_z$ vertical bins (i.e., in the $z$ direction that is perpendicular to the planes), and count the number of sphere centers that fall into each bin at different height $z$, respectively. Here we choose $N_z$ to be 50 such that the results do not vary sensitively upon perturbing the thickness of the bins, and do not lose local information as well. In Fig. \ref{fig_5} we plot the density profiles $\rho(z)$ of monodisperse (i.e., equal-sized) hard-sphere packings at various $H$ as examples. For all values of $H$ shown in Fig. \ref{fig_5}, there are two major peaks adjacent to the two hard walls corresponding to the two contacting layers and a few smaller peaks in the interior of the packings, whose intensities decrease with increasing $H$. This indicates increasing disorder as $H$ increases, which is due to decreasing confinement frustration.

\begin{figure}[H]
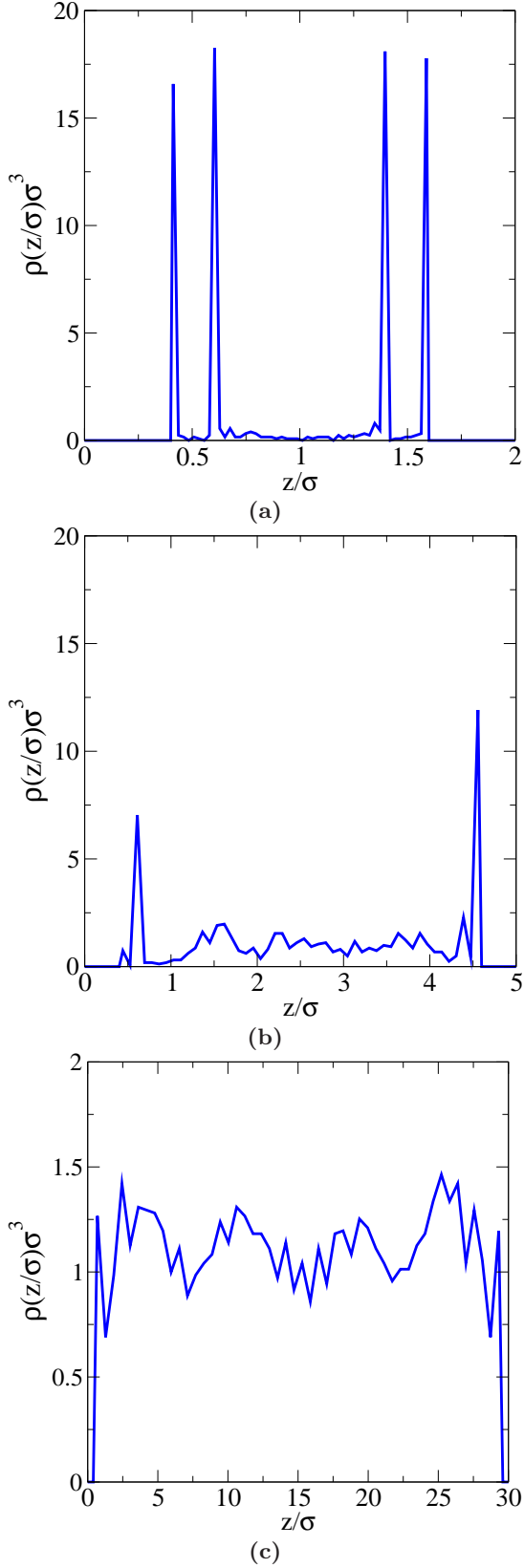

\begin{center}
$\begin{array}{c}\\
\includegraphics[width=0.40\textwidth]{fig6a.eps} \\
\mbox{\bf (a)} \\
\includegraphics[width=0.40\textwidth]{fig6b.eps} \\
\mbox{\bf (b)} \\
\includegraphics[width=0.40\textwidth]{fig6c.eps} \\
\mbox{\bf (c)} \\
\end{array}$
\end{center}
\caption{(Color online) Density profiles of confined binary hard-sphere packings at various $H/\sigma$ at $\alpha = 2/3$ and $x = 0.5$, where $\sigma$ is the average sphere diameter. (a) $H/\sigma = 2.0$. (b) $H/\sigma = 5.0$. (c) $H/\sigma = 30.0$.} \label{fig_6}
\end{figure}

In Fig. \ref{fig_6} we plot the density profiles $\rho(z)$ of \textit{binary} hard-sphere packings at $\alpha = 2/3$ and $x = 0.5$ for certain values of $H$. Similar behaviors could be observed in these binary systems as in their monodisperse counterparts. Interestingly, we can observe two split peaks of contacting large and small spheres adjacent to the hard walls. We note that at the same $H$ the density profiles of the binary systems appear to be flatter than those of the monodisperse systems, implying increasing disorder due to the aforementioned size-disparity frustration. We will be able to quantify the order of the system in more details as we introduce an order metric later. It is noteworthy that from Figs. \ref{fig_5} and \ref{fig_6}, we can see that when the plane separation distance is of the order of two large-sphere diameters or less, the packings possess layered structures, which are salient 2D features; when the plane separation distance exceeds about 30 large-sphere diameters, the packings approach 3D bulk packings and exhibit relatively flat density profiles, which is due to the decreasing confinement frustration.

\begin{figure}[H]
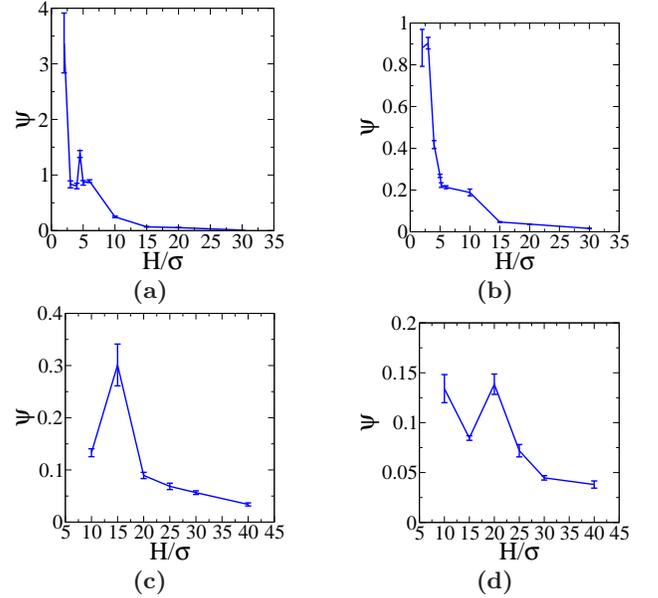

\begin{center}
$\begin{array}{c@{\hspace{1.0cm}}c}\\
\includegraphics[width=0.20\textwidth]{fig7a.eps} &
\includegraphics[width=0.20\textwidth]{fig7b.eps} \\
\mbox{\bf (a)} & \mbox{\bf (b)} \\
\includegraphics[width=0.20\textwidth]{fig7c.eps} &
\includegraphics[width=0.20\textwidth]{fig7d.eps} \\
\mbox{\bf (c)} & \mbox{\bf (d)}
\end{array}$
\end{center}
\caption{(Color online) Plot of the order metric $\psi$ as a function of $H/\sigma$ obtained at various $\alpha$ and $x$, where $\sigma$ is the average sphere diameter. Each data point in the plot is obtained by averaging over 10 packings (vertical bars represent one standard deviation). (a) $\alpha = 1.0$ (equal-sized spheres). (b) $\alpha = 2/3, x = 0.5$. (c) $\alpha = 0.2, x = 0.95$. (d) $\alpha = 0.2, x = 0.97$.} \label{fig_7}
\end{figure}

Based on density fluctuations in the direction perpendicular to the hard walls, we define an order metric $\psi$ to quantify the order of packings:
\begin{equation}
\label{eq_7} \psi = \frac{1}{H-\sigma_L-2\delta}\int_{\sigma_L/2+\delta}^{H-\sigma_L/2-\delta}\frac{(\rho(z)-{\bar \rho})^2}{{\bar \rho}^2}\mathrm{d}z,
\end{equation}
where $\sigma_L$ is the diameter of the large sphere, $\delta$ is chosen as a small quantity (relative to $\sigma_L$) to exclude the layers in contact with the two hard walls, $\rho(z)$ is the particle density at height $z$, and ${\bar \rho}$ is the particle density averaged over different heights $z$ excluding the layers in contact with the hard walls. In Fig. \ref{fig_7} we plot the computed $\psi$ as a function of $H$ at different $\alpha$ and $x$. As one can see, the packings generally become more disordered, as $H$ increases due to decreasing confinement frustration. There are local maxima and minima in $\psi$ as $H$ varies across certain critical values at certain values of $\alpha$ and $x$, which is due to aforementioned discontinuous transitions.

\begin{figure}[H]
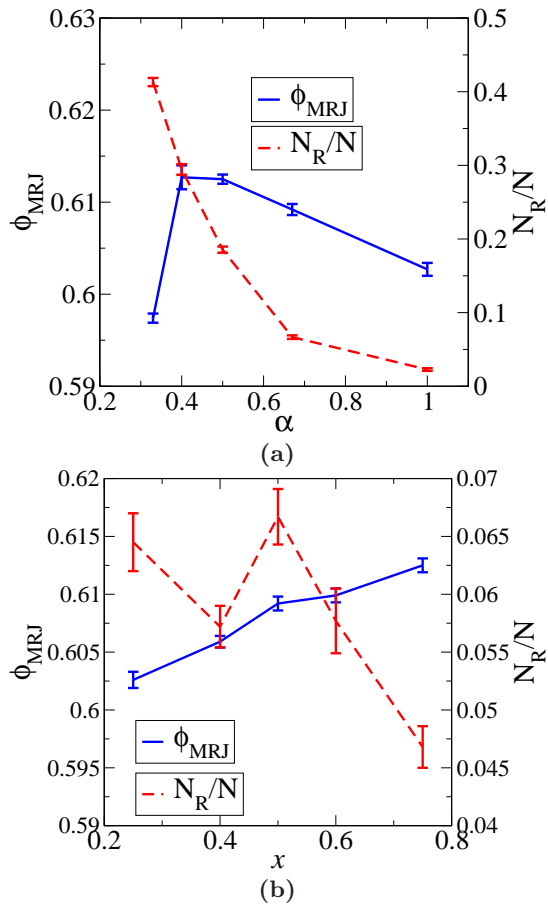

\begin{center}
$\begin{array}{c}\\
\includegraphics[width=0.40\textwidth]{fig8a.eps} \\
\mbox{\bf (a)} \\
\includegraphics[width=0.40\textwidth]{fig8b.eps} \\
\mbox{\bf (b)} \\
\end{array}$
\end{center}
\caption{(Color online) (a) Plot of the average packing fraction $\phi_{\mbox{\scriptsize{MRJ}}}$ and rattler fraction $N_R/N$ as functions of $\alpha$ obtained at $H/\sigma = 5.0$ and $x = 0.5$, where $\sigma$ is the average sphere diameter. Each data point in the plot is obtained by averaging over 10 packings (vertical bars represent one standard deviation). (b) Plot of the average packing fraction $\phi_{\mbox{\scriptsize{MRJ}}}$ and rattler fraction $N_R/N$ as functions of $x$ obtained at $H/\sigma = 5.0$ and $\alpha = 2/3$, where $\sigma$ is the average sphere diameter. Each data point in the plot is obtained by averaging over 10 packings (vertical bars represent one standard deviation).} \label{fig_8}
\end{figure}

After looking at various statistics as $H$ varies at different $\alpha$ and $x$, we investigate the effects of size contrast and composition on the packings at given $H$ in details. In Fig. \ref{fig_8}(a), we plot $\phi_{\mbox{\scriptsize{MRJ}}}$ and rattler fraction $N_R/N$ as functions of $\alpha$ at $H/\sigma = 5.0$ and $x = 0.5$. We find that $\phi_{\mbox{\scriptsize{MRJ}}}$ reaches a maximum at about $\alpha = 0.4$, which is related to the fact that at such $\alpha$, most small spheres participate in the backbone and fill the interstices between the large spheres, leading to efficient packings. However, $N_R/N$ decreases monotonically with $\alpha$ as size-disparity frustration decreases. This trends are similar to those observed in bulk counterparts \cite{Ho13}. In Fig. \ref{fig_8}(b), we plot $\phi_{\mbox{\scriptsize{MRJ}}}$ and $N_R/N$ as functions of $x$ at $H/\sigma = 5.0$ and $\alpha = 2/3$. We find that in the range of $x$ studied, $\phi_{\mbox{\scriptsize{MRJ}}}$ increases monotonically as $x$ increases since more small spheres are available to fill the small ``voids'' left by the large spheres, and $N_R/N$ reaches a maximum near $x = 0.5$ due to maximized size-disparity frustration, as mentioned earlier. Note that if $x$ were to increase beyond the range currently investigated, we would expect $\phi_{\mbox{\scriptsize{MRJ}}}$ to eventually reach a maximum, similar to the behaviors reported in previous studies of other binary systems \cite{Ho12, Bi09, Da10}, though this remains to be verified by future simulations.

\begin{figure}[H]
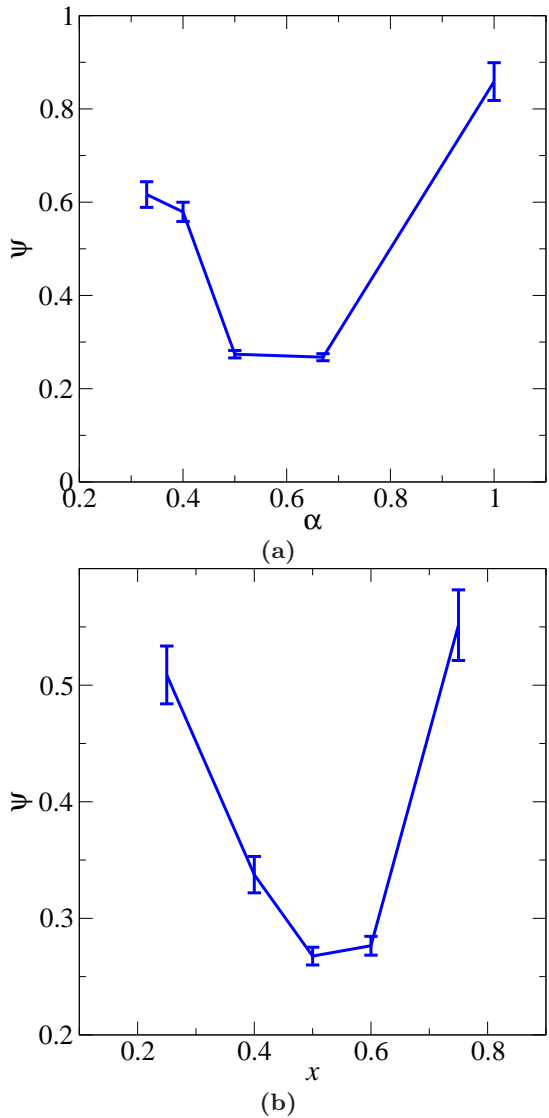

\begin{center}
$\begin{array}{c}\\
\includegraphics[width=0.40\textwidth]{fig9a.eps} \\
\mbox{\bf (a)} \\
\includegraphics[width=0.40\textwidth]{fig9b.eps} \\
\mbox{\bf (b)} \\
\end{array}$
\end{center}
\caption{(Color online) (a) Plot of the order metric $\psi$ as a function of $\alpha$ obtained at $H/\sigma = 5.0$ and $x = 0.5$, where $\sigma$ is the average sphere diameter. Each data point in the plot is obtained by averaging over 10 packings (vertical bars represent one standard deviation). (b) Plot of the order metric $\psi$ as a function of $x$ obtained at $H/\sigma = 5.0$ and $\alpha = 2/3$, where $\sigma$ is the average sphere diameter. Each data point in the plot is obtained by averaging over 10 packings (vertical bars represent one standard deviation).} \label{fig_9}
\end{figure}

Furthermore, we compute and plot in Fig. \ref{fig_9}(a) the order metric $\psi$ defined in Eq. \ref{eq_7} as a function of $\alpha$ at $H/\sigma = 5.0$ and $x = 0.5$. We find that in this case the minimum of $\psi$ occurs at about $\alpha = 2/3$ as $\alpha$ varies. This is because at small $\alpha$, large spheres possess relatively large exclusion volumes empty of other sphere centers, leading to large fluctuations in $\rho(z)$; while at large $\alpha$, layered structures are favored, which possess large density fluctuations in the direction perpendicular to the planes as well. In Fig. \ref{fig_9}(b), we plot $\psi$ as a function of $x$ at $H/\sigma = 5.0$ and $\alpha = 2/3$. We find that the minimum of $\psi$ occurs at about $x = 0.5$ as $x$ varies, implying maximized disorder due to favorable mixing entropy.

\begin{figure}[H]
\begin{center}
$\begin{array}{c@{\hspace{1.0cm}}c}\\
\includegraphics[width=0.22\textwidth]{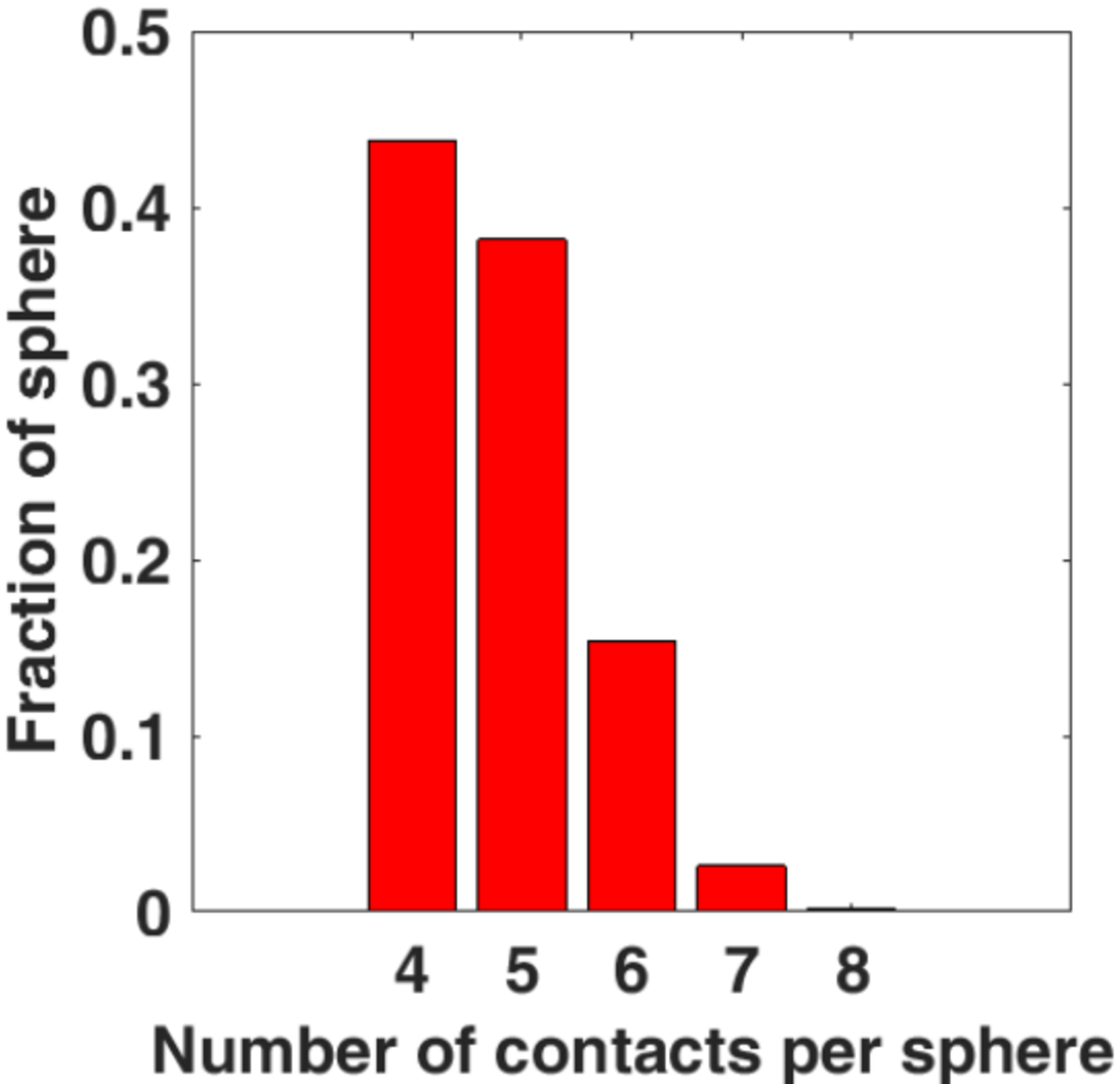} &
\includegraphics[width=0.22\textwidth]{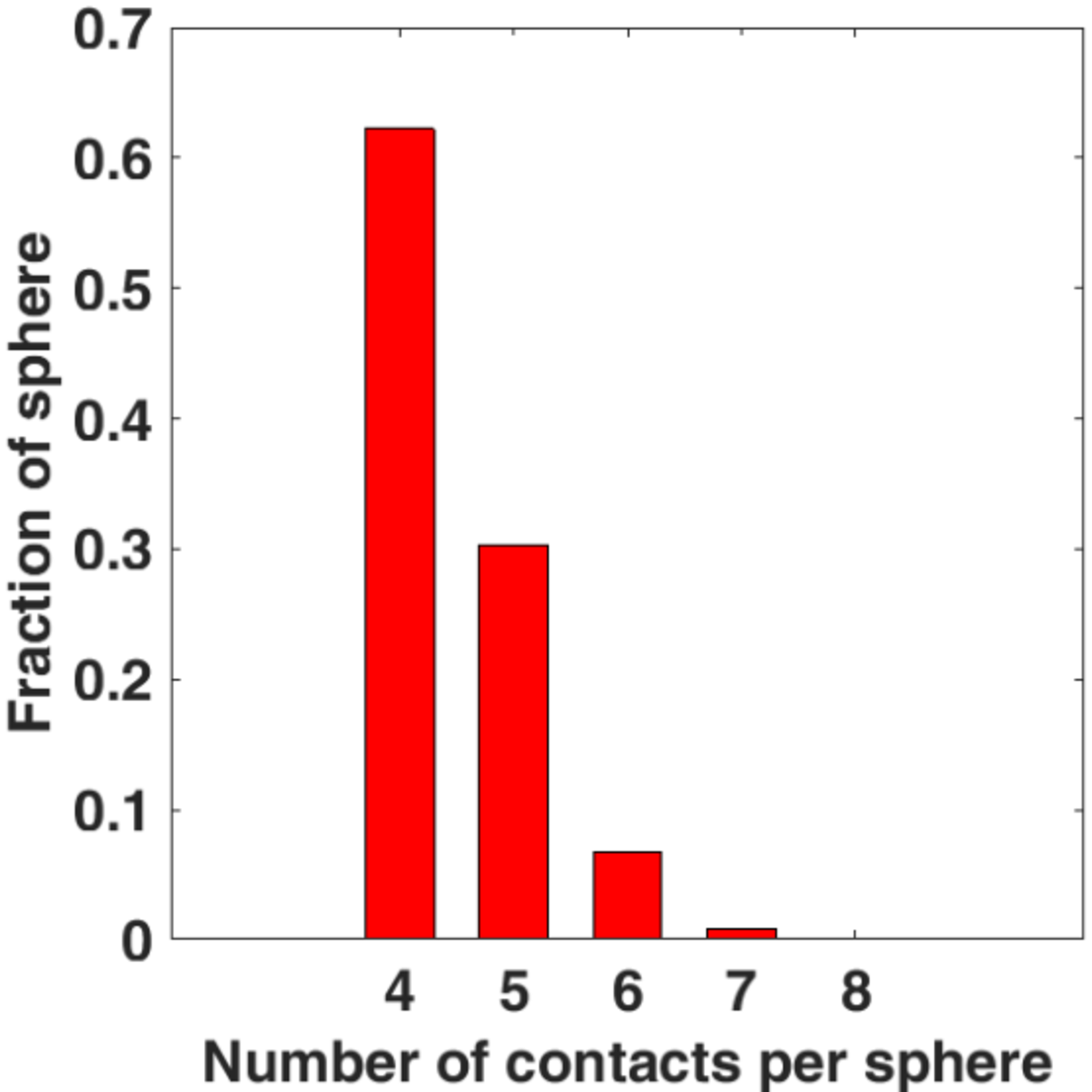} \\
\mbox{\bf (a)} & \mbox{\bf (b)} \\
\includegraphics[width=0.22\textwidth]{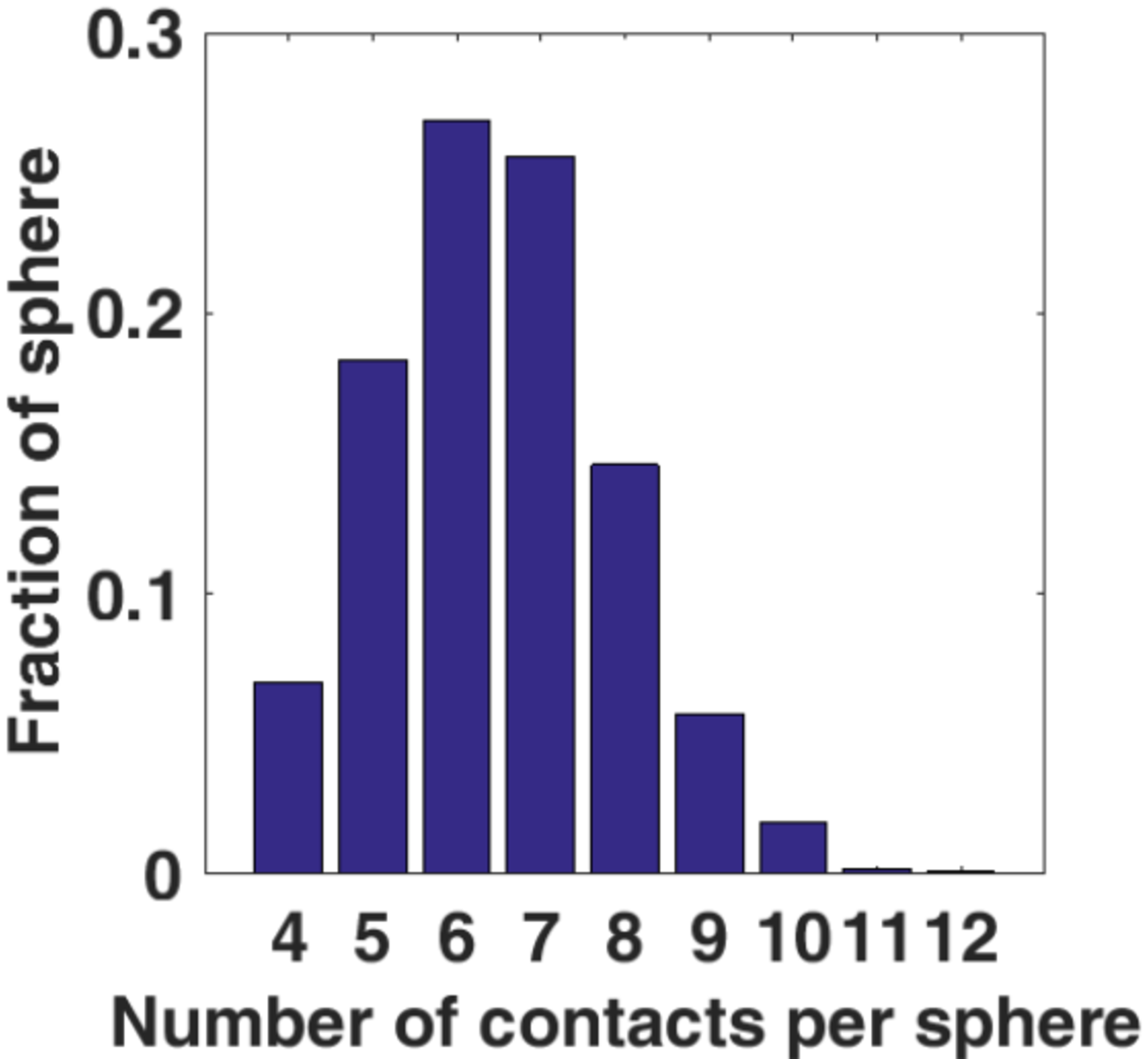} &
\includegraphics[width=0.22\textwidth]{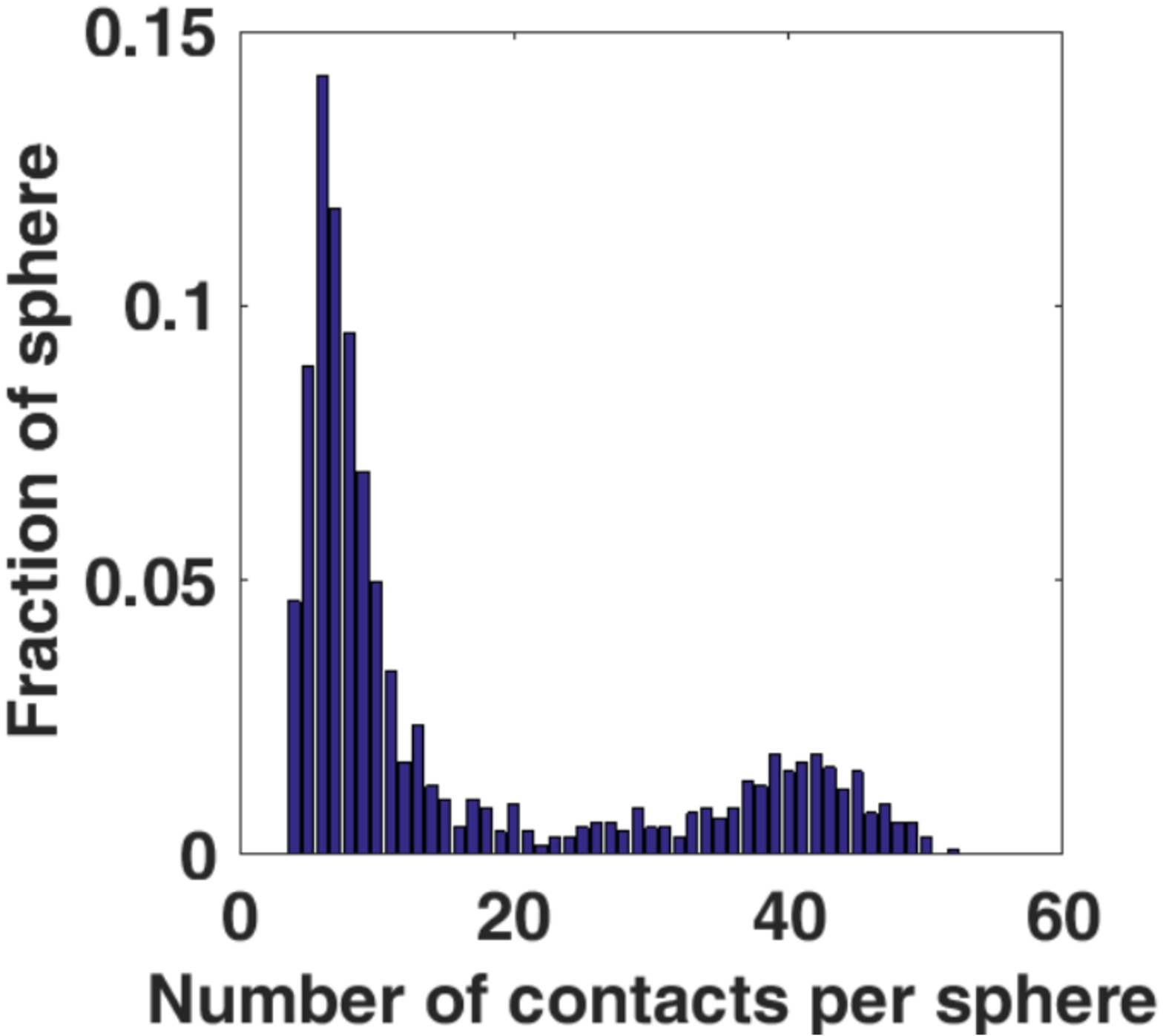} \\
\mbox{\bf (c)} & \mbox{\bf (d)}
\end{array}$
\end{center}
\caption{(Color online) Histograms depicting the average fractions of jammed spheres with specified numbers of contacts for $H/\sigma = 5.0$, $\alpha = 2/3 $, and $x = 0.5$ [(a) and (c)], and $H/\sigma = 20.0$, $\alpha = 0.2 $ and $x = 0.97$ [(b) and (d)], where $\sigma$ is the average sphere diameter. Plots (a), and (b) each represent the average over 10 packings of the fractions of jammed small spheres with specified contact numbers, and plots (c), and (d) each represent average fraction for the jammed large spheres.} \label{fig_10}
\end{figure}

Subsequently, we employ more structural descriptors to further characterize the packings. In Fig. \ref{fig_10} we plot the detailed contact distributions of backbone spheres for two representative cases: $H/\sigma = 5.0$, $\alpha = 2/3 $, and $x = 0.5$, and $H/\sigma = 20.0$, $\alpha = 0.2$ and $x = 0.97$. We find that in both cases most jammed small spheres have 4 particle-particle and particle-plane contacts, similar to the corresponding bulk cases \cite{Ho13}. However, a small percentage of small spheres possess more than 6 contacts and there are two separate peaks in the contact distribution of jammed large spheres at $H/\sigma = 20.0$, $\alpha = 2/3$, and $x = 0.5$, which are different from the bulk cases \cite{Ho13}. Also, the percentages of jammed small and large spheres that possess few contacts in these confined packings are higher than their bulk counterparts \cite{Ho13}, which decrease with $H$. These phenomena are caused by the aforementioned confinement frustration, which modify the local arrangements near the hard walls.

\begin{figure}[H]
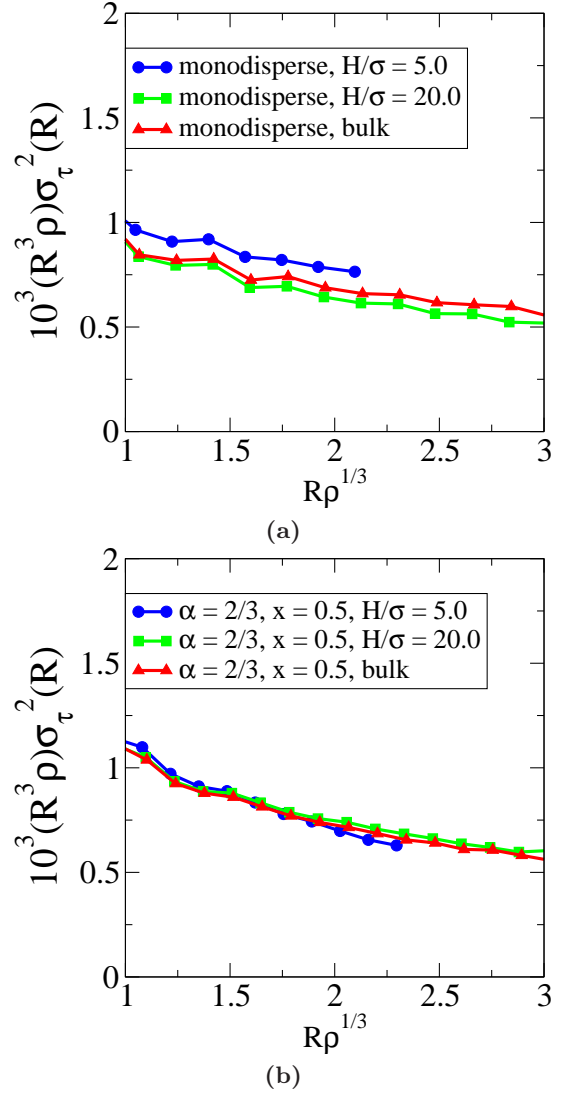

\begin{center}
$\begin{array}{c}\\
\includegraphics[width=0.40\textwidth]{fig11a.eps} \\
\mbox{\bf (a)} \\
\includegraphics[width=0.40\textwidth]{fig11b.eps} \\
\mbox{\bf (b)} \\
\end{array}$
\end{center}
\caption{(Color online) Scaled local volume-fraction variance $10^3(R^3\rho)\sigma_{\tau}^2(R)$ as function of the window radius $R$ for confined hard-sphere packings in (a) monodisperse case and (b) binary case of $\alpha = 2/3$, and $x = 0.5$ at $H/\sigma=5.0, 20.0$ and their bulk counterparts. Note that $10^3(R^3\rho)\sigma_{\tau}^2(R)$ tends to zero as $R$ increases without bound, i.e., $\sigma_{\tau}^2(R)$ decays faster than $R^{-3}$, implying hyperuniformity \cite{Za09, Za11a, Za11b}. Moreover, these confined packings possess essentially the same local volume-fraction variance $\sigma_{\tau}^2(R)$ as their bulk counterparts, which reflect essentially the same level of hyperuniformity.} \label{fig_11}
\end{figure}

The local volume-fraction variance $\sigma_{\tau}^2(R)$ \cite{Lu90, Za09} in a spherical observation window has been employed in our previous work to characterize bulk packings \cite{Lu90, Za09, Za11a, Za11b, Za11c, Dr15}. Here we extend the application of this descriptor to confined packings, with the constraint that the radius of the window $R$ should not exceed half of the plane separation distance $H$, i.e., $R \leq H/2$. The volume-fraction variance is sampled by randomly placing observation windows in the system under the constraint that the windows should be entirely within the confining space. We also compute $\sigma_{\tau}^2(R)$ for the bulk counterparts of these confined packings for comparison. Note that if a system is hyperuniform, its local volume-fraction variance $\sigma_{\tau}^2(R)$ should decay faster than $R^{-d}$ (where $d$ is the Euclidean space dimension); i.e., in three dimensions the scaled local volume-fraction variance $R^3\sigma_{\tau}^2(R)$ should tend to zero as $R$ approaches infinity \cite{Za09, Za11a, Za11b}. We find that these packings possess essentially the same local volume-fraction variance $\sigma_{\tau}^2(R)$ as a function of $R$ as that of their bulk counterparts, which reflect effectively the same level of hyperuniformity, as shown in Fig. \ref{fig_11} \footnote{This is not surprising since $\sigma_{\tau}^2(R)$ effectively homogenizes information at different heights between the two planes, which leads to its insensitivity to confinement.}. Note that a disordered hyperuniform system is an exotic amorphous state of matter whose local volume-fraction fluctuations asymptotically decay faster than the reciprocal of the volume of the observation window, which the decay associated with typical disordered systems. The hyperuniform decay rate implies that the local volume fraction approaches the global value $\phi_{\mbox{\scriptsize{MRJ}}}$ anomalously fast \cite{Za09, Za11a, Za11b, Za11c, Dr15}. Nonetheless, these results for $\sigma_{\tau}^2(R)$ demonstrate that hyperuniformity appears to be a signature of MRJ packings, whether they exist in the bulk or under confinement.

\section{Conclusions}
In this paper, we generalized the TJ sequential linear algorithm to generate exactly isostatic putative MRJ binary hard-sphere packings confined between two parallel hard planes over a large range of plane separation distances $H$, small to large sphere radius ratio $\alpha$ and small sphere relative concentration $x$. We observe that these confined packings generally possess structural characteristics that are distinctly different from their bulk counterparts, including lower packing fractions, and higher rattler fractions, as well as varying degrees of disorder and particle contacts. This is due to what we call confinement frustration. We find that an order metric $\psi$, which is based on the number density fluctuations in the direction perpendicular to the hard walls, is a useful measure of the degree of order (disorder). By employing the local volume-fraction variance $\sigma_{\tau}^2(R)$, we find that these packings possess essentially the same level of hyperuniformity as their bulk counterparts. We also observe that the packing characteristics depend sensitively on $H$, $\alpha$ and $x$ for $H$ smaller than 30 large sphere diameters due to the effects of confinement frustration, size-disparity frustration and discontinuous transitions. We have also observed that the packings gradually transition from ones with 2D-like layered structures to 3D bulk systems as $H$ increases from two to 30 large sphere diameters.

The dependence of packing density on the confinement size has important practical implications in various powder technologies, where density is crucial to material properties and fabrication cost \cite{Ko01, Mi81, Ki76}. By looking at how the packing fraction changes with plane separation distance $H$ in our simulation, we acquire knowledge about how the thickness of depositing layers affect the density and associated mechanical and transport properties of the resulting structures given the starting materials in powder technologies. In particular, we find that small-sized particles are favored to suppress the packing inefficiency caused by boundary and finite-size effects and they should be used to guarantee high density and superior material performances. When producing structures that have a dimension comparable to particle sizes, extremely large percentage of small particles should be mixed with a tiny percentage of large particles to suppress the density variance of individual samples caused by small errors in the thickness of depositing layers. Also, our findings could be potentially useful in battery applications, e.g. solid oxide fuel cell electrode materials. When the length scale of the shortest dimension is less than about 30 particle diameters, boundary, and finite-size effects should be taken into account when we evaluate and tailor the macroscopic properties and performance of the electrode materials. In future work, we will investigate the optimal thickness, particle size ratio and composition that optimize the electrochemical properties of electrode materials such as maximizing reaction rate and ionic and electronic conductivities. In addition, we might be able to design novel photonic structures and devices by tailoring the size distribution of nanoparticles in finite particle packings \cite{Pr04} according to information acquired in our simulations.

Besides the packings of binary hard spheres confined between two parallel hard planes we have studied in this work, there are many other interesting packing problems in confined space that remain to be investigated \cite{Sh12, Fo15, Mi11}. For example, while \textit{bulk} MRJ packings of nonspherical particles that span a wide range of shapes, including ellipsoids \cite{Do04b, Ma05, Sc15}, superballs \cite{Ji10, Ng14}, and polyhedra \cite{Ji11a, Ch14a, Za11c, Ba13} have been studied in detail, nothing is known about confined MRJ packings of nonspherical particles. Understanding how confined MRJ packings of nonspherical particles differ from their bulk counterparts and confined MRJ sphere packings are outstanding questions. Extensions of this work to DNA packaging is an interesting avenue for future research \cite{Mi11}.

\begin{acknowledgments}
We are grateful to Steven Atkinson and Adam B. Hopkins for very helpful discussions. This work was supported by the National Science Foundation under Grant No. DMS-1211087.
\end{acknowledgments}

%


\begin{thebibliography}{82}%
\makeatletter
\providecommand \@ifxundefined [1]{%
 \@ifx{#1\undefined}
}%
\providecommand \@ifnum [1]{%
 \ifnum #1\expandafter \@firstoftwo
 \else \expandafter \@secondoftwo
 \fi
}%
\providecommand \@ifx [1]{%
 \ifx #1\expandafter \@firstoftwo
 \else \expandafter \@secondoftwo
 \fi
}%
\providecommand \natexlab [1]{#1}%
\providecommand \enquote  [1]{``#1''}%
\providecommand \bibnamefont  [1]{#1}%
\providecommand \bibfnamefont [1]{#1}%
\providecommand \citenamefont [1]{#1}%
\providecommand \href@noop [0]{\@secondoftwo}%
\providecommand \href [0]{\begingroup \@sanitize@url \@href}%
\providecommand \@href[1]{\@@startlink{#1}\@@href}%
\providecommand \@@href[1]{\endgroup#1\@@endlink}%
\providecommand \@sanitize@url [0]{\catcode `\\12\catcode `\$12\catcode
  `\&12\catcode `\#12\catcode `\^12\catcode `\_12\catcode `\%12\relax}%
\providecommand \@@startlink[1]{}%
\providecommand \@@endlink[0]{}%
\providecommand \url  [0]{\begingroup\@sanitize@url \@url }%
\providecommand \@url [1]{\endgroup\@href {#1}{\urlprefix }}%
\providecommand \urlprefix  [0]{URL }%
\providecommand \Eprint [0]{\href }%
\providecommand \doibase [0]{http://dx.doi.org/}%
\providecommand \selectlanguage [0]{\@gobble}%
\providecommand \bibinfo  [0]{\@secondoftwo}%
\providecommand \bibfield  [0]{\@secondoftwo}%
\providecommand \translation [1]{[#1]}%
\providecommand \BibitemOpen [0]{}%
\providecommand \bibitemStop [0]{}%
\providecommand \bibitemNoStop [0]{.\EOS\space}%
\providecommand \EOS [0]{\spacefactor3000\relax}%
\providecommand \BibitemShut  [1]{\csname bibitem#1\endcsname}%
\let\auto@bib@innerbib\@empty
\bibitem [{\citenamefont {Frenkel}\ and\ \citenamefont {Smit}(2001)}]{Fr01}%
  \BibitemOpen
  \bibfield  {author} {\bibinfo {author} {\bibfnamefont {D.}~\bibnamefont
  {Frenkel}}\ and\ \bibinfo {author} {\bibfnamefont {B.}~\bibnamefont {Smit}},\
  }\href@noop {} {\emph {\bibinfo {title} {Understanding Molecular Simulation:
  From Algorithms to Applications}}},\ Vol.~\bibinfo {volume} {1}\ (\bibinfo
  {publisher} {Academic Press, San Diego},\ \bibinfo {year}
  {2001})\BibitemShut {NoStop}%
\bibitem [{\citenamefont {Jadrich}\ and\ \citenamefont
  {Schweizer}(2013{\natexlab{a}})}]{Ja13a}%
  \BibitemOpen
  \bibfield  {author} {\bibinfo {author} {\bibfnamefont {R.}~\bibnamefont
  {Jadrich}}\ and\ \bibinfo {author} {\bibfnamefont {K.~S.}\ \bibnamefont
  {Schweizer}},\ }\href@noop {} {\bibfield  {journal} {\bibinfo  {journal} {J.
  Chem. Phys.}\ }\textbf {\bibinfo {volume} {139}},\ \bibinfo {pages} {054501}
  (\bibinfo {year} {2013}{\natexlab{a}})}\BibitemShut {NoStop}%
\bibitem [{\citenamefont {Jadrich}\ and\ \citenamefont
  {Schweizer}(2013{\natexlab{b}})}]{Ja13b}%
  \BibitemOpen
  \bibfield  {author} {\bibinfo {author} {\bibfnamefont {R.}~\bibnamefont
  {Jadrich}}\ and\ \bibinfo {author} {\bibfnamefont {K.~S.}\ \bibnamefont
  {Schweizer}},\ }\href@noop {} {\bibfield  {journal} {\bibinfo  {journal} {J.
  Chem. Phys.}\ }\textbf {\bibinfo {volume} {139}},\ \bibinfo {pages} {054502}
  (\bibinfo {year} {2013}{\natexlab{b}})}\BibitemShut {NoStop}%
\bibitem [{\citenamefont {Chaikin}\ and\ \citenamefont
  {Lubensky}(2000)}]{Ch00}%
  \BibitemOpen
  \bibfield  {author} {\bibinfo {author} {\bibfnamefont {P.~M.}\ \bibnamefont
  {Chaikin}}\ and\ \bibinfo {author} {\bibfnamefont {T.~C.}\ \bibnamefont
  {Lubensky}},\ }\href@noop {} {\emph {\bibinfo {title} {Principles of
  Condensed Matter Physics}}},\ Vol.~\bibinfo {volume} {1}\ (\bibinfo
  {publisher} {Cambridge University Press, New York},\ \bibinfo {year}
  {2000})\BibitemShut {NoStop}%
\bibitem [{\citenamefont {Zallen}(1983)}]{Za83}%
  \BibitemOpen
  \bibfield  {author} {\bibinfo {author} {\bibfnamefont {R.}~\bibnamefont
  {Zallen}},\ }\href@noop {} {\emph {\bibinfo {title} {The Physics of Amorphous
  Solids}}}\ (\bibinfo  {publisher} {Wiley, New York},\ \bibinfo {year}
  {1983})\BibitemShut {NoStop}%
\bibitem [{\citenamefont {Torquato}(2002)}]{To02}%
  \BibitemOpen
  \bibfield  {author} {\bibinfo {author} {\bibfnamefont {S.}~\bibnamefont
  {Torquato}},\ }\href@noop {} {\emph {\bibinfo {title} {Random Heterogeneous
  Materials: Microstructure and Macroscopic Properties}}},\ Vol.~\bibinfo
  {volume} {16}\ (\bibinfo  {publisher} {Springer: New York},\ \bibinfo {year}
  {2002})\BibitemShut {NoStop}%
\bibitem [{\citenamefont {de~Gennes}(1999)}]{De99}%
  \BibitemOpen
  \bibfield  {author} {\bibinfo {author} {\bibfnamefont {P.~G.}\ \bibnamefont
  {de~Gennes}},\ }\href@noop {} {\bibfield  {journal} {\bibinfo  {journal}
  {Rev. Mod. Phys.}\ }\textbf {\bibinfo {volume} {71}},\ \bibinfo {pages}
  {S374} (\bibinfo {year} {1999})}\BibitemShut {NoStop}%
\bibitem [{\citenamefont {Nesterenko}(2001)}]{Ne01}%
  \BibitemOpen
  \bibfield  {author} {\bibinfo {author} {\bibfnamefont {V.~F.}\ \bibnamefont
  {Nesterenko}},\ }\href@noop {} {\emph {\bibinfo {title} {Dynamics of
  Heterogeneous Materials}}}\ (\bibinfo  {publisher} {Springer, New York},\
  \bibinfo {year} {2001})\BibitemShut {NoStop}%
\bibitem [{\citenamefont {Sahimi}(2003{\natexlab{a}})}]{Sa03a}%
  \BibitemOpen
  \bibfield  {author} {\bibinfo {author} {\bibfnamefont {M.}~\bibnamefont
  {Sahimi}},\ }\href@noop {} {\emph {\bibinfo {title} {Heterogeneous Materials
  I: Linear Transport and Optical Properties}}},\ Vol.~\bibinfo {volume} {1}\
  (\bibinfo  {publisher} {Springer, New York},\ \bibinfo {year}
  {2003})\BibitemShut {NoStop}%
\bibitem [{\citenamefont {Sahimi}(2003{\natexlab{b}})}]{Sa03b}%
  \BibitemOpen
  \bibfield  {author} {\bibinfo {author} {\bibfnamefont {M.}~\bibnamefont
  {Sahimi}},\ }\href@noop {} {\emph {\bibinfo {title} {Heterogeneous Materials
  II: Nonlinear and Breakdown Properties and Atomistic Modeling}}},\
  Vol.~\bibinfo {volume} {2}\ (\bibinfo  {publisher} {Springer, New York},\
  \bibinfo {year} {2003})\BibitemShut {NoStop}%
\bibitem [{\citenamefont {Olmos}\ \emph {et~al.}(2009)\citenamefont {Olmos},
  \citenamefont {Martin},\ and\ \citenamefont {Bouvard}}]{Ol09}%
  \BibitemOpen
  \bibfield  {author} {\bibinfo {author} {\bibfnamefont {L.}~\bibnamefont
  {Olmos}}, \bibinfo {author} {\bibfnamefont {C.~L.}\ \bibnamefont {Martin}}, \
  and\ \bibinfo {author} {\bibfnamefont {D.}~\bibnamefont {Bouvard}},\
  }\href@noop {} {\bibfield  {journal} {\bibinfo  {journal} {Powder Technol.}\
  }\textbf {\bibinfo {volume} {190}},\ \bibinfo {pages} {134} (\bibinfo {year}
  {2009})}\BibitemShut {NoStop}%
\bibitem [{\citenamefont {Zohdi}(2014{\natexlab{a}})}]{Zo14a}%
  \BibitemOpen
  \bibfield  {author} {\bibinfo {author} {\bibfnamefont {T.~I.}\ \bibnamefont
  {Zohdi}},\ }\href@noop {} {\bibfield  {journal} {\bibinfo  {journal} {Math.
  Mech. Solids}\ }\textbf {\bibinfo {volume} {19}},\ \bibinfo {pages} {93}
  (\bibinfo {year} {2014}{\natexlab{a}})}\BibitemShut {NoStop}%
\bibitem [{\citenamefont {Zohdi}(2014{\natexlab{b}})}]{Zo14b}%
  \BibitemOpen
  \bibfield  {author} {\bibinfo {author} {\bibfnamefont {T.~I.}\ \bibnamefont
  {Zohdi}},\ }\href@noop {} {\bibfield  {journal} {\bibinfo  {journal} {Comput.
  Mech.}\ }\textbf {\bibinfo {volume} {54}},\ \bibinfo {pages} {171} (\bibinfo
  {year} {2014}{\natexlab{b}})}\BibitemShut {NoStop}%
\bibitem [{\citenamefont {Hales}(2005)}]{Ha05}%
  \BibitemOpen
  \bibfield  {author} {\bibinfo {author} {\bibfnamefont {T.~C.}\ \bibnamefont
  {Hales}},\ }\href@noop {} {\bibfield  {journal} {\bibinfo  {journal} {Ann.
  Math.}\ }\textbf {\bibinfo {volume} {162}},\ \bibinfo {pages} {1065}
  (\bibinfo {year} {2005})}\BibitemShut {NoStop}%
\bibitem [{\citenamefont {Hansen}\ and\ \citenamefont {McDonald}(1986)}]{Ha86}%
  \BibitemOpen
  \bibfield  {author} {\bibinfo {author} {\bibfnamefont {J.-P.}\ \bibnamefont
  {Hansen}}\ and\ \bibinfo {author} {\bibfnamefont {I.~R.}\ \bibnamefont
  {McDonald}},\ }\href@noop {} {\emph {\bibinfo {title} {Theory of Simple
  Liquids}}}\ (\bibinfo  {publisher} {Academic Press, London},\ \bibinfo {year}
  {1986})\BibitemShut {NoStop}%
\bibitem [{\citenamefont {Torquato}\ \emph {et~al.}(2000)\citenamefont
  {Torquato}, \citenamefont {Truskett},\ and\ \citenamefont
  {Debenedetti}}]{To00}%
  \BibitemOpen
  \bibfield  {author} {\bibinfo {author} {\bibfnamefont {S.}~\bibnamefont
  {Torquato}}, \bibinfo {author} {\bibfnamefont {T.~M.}\ \bibnamefont
  {Truskett}}, \ and\ \bibinfo {author} {\bibfnamefont {P.~G.}\ \bibnamefont
  {Debenedetti}},\ }\href@noop {} {\bibfield  {journal} {\bibinfo  {journal}
  {Phys. Rev. Lett.}\ }\textbf {\bibinfo {volume} {84}},\ \bibinfo {pages}
  {2064} (\bibinfo {year} {2000})}\BibitemShut {NoStop}%
\bibitem [{\citenamefont {Torquato}\ and\ \citenamefont
  {Stillinger}(2010)}]{To10a}%
  \BibitemOpen
  \bibfield  {author} {\bibinfo {author} {\bibfnamefont {S.}~\bibnamefont
  {Torquato}}\ and\ \bibinfo {author} {\bibfnamefont {F.~H.}\ \bibnamefont
  {Stillinger}},\ }\href@noop {} {\bibfield  {journal} {\bibinfo  {journal}
  {Rev. Mod. Phys.}\ }\textbf {\bibinfo {volume} {82}},\ \bibinfo {pages}
  {2633} (\bibinfo {year} {2010})}\BibitemShut {NoStop}%
\bibitem [{\citenamefont {Rintoul}\ and\ \citenamefont
  {Torquato}(1996)}]{Ri96}%
  \BibitemOpen
  \bibfield  {author} {\bibinfo {author} {\bibfnamefont {M.~D.}\ \bibnamefont
  {Rintoul}}\ and\ \bibinfo {author} {\bibfnamefont {S.}~\bibnamefont
  {Torquato}},\ }\href@noop {} {\bibfield  {journal} {\bibinfo  {journal} {J.
  Chem. Phys.}\ }\textbf {\bibinfo {volume} {105}},\ \bibinfo {pages} {9258}
  (\bibinfo {year} {1996})}\BibitemShut {NoStop}%
\bibitem [{\citenamefont {Torquato}\ and\ \citenamefont {Jiao}(2010)}]{To10b}%
  \BibitemOpen
  \bibfield  {author} {\bibinfo {author} {\bibfnamefont {S.}~\bibnamefont
  {Torquato}}\ and\ \bibinfo {author} {\bibfnamefont {Y.}~\bibnamefont
  {Jiao}},\ }\href@noop {} {\bibfield  {journal} {\bibinfo  {journal} {Phys.
  Rev. E}\ }\textbf {\bibinfo {volume} {82}},\ \bibinfo {pages} {061302}
  (\bibinfo {year} {2010})}\BibitemShut {NoStop}%
\bibitem [{\citenamefont {Atkinson}\ \emph {et~al.}(2013)\citenamefont
  {Atkinson}, \citenamefont {Stillinger},\ and\ \citenamefont
  {Torquato}}]{At13}%
  \BibitemOpen
  \bibfield  {author} {\bibinfo {author} {\bibfnamefont {S.}~\bibnamefont
  {Atkinson}}, \bibinfo {author} {\bibfnamefont {F.~H.}\ \bibnamefont
  {Stillinger}}, \ and\ \bibinfo {author} {\bibfnamefont {S.}~\bibnamefont
  {Torquato}},\ }\href@noop {} {\bibfield  {journal} {\bibinfo  {journal}
  {Phys. Rev. E}\ }\textbf {\bibinfo {volume} {88}},\ \bibinfo {pages} {062208}
  (\bibinfo {year} {2013})}\BibitemShut {NoStop}%
\bibitem [{\citenamefont {Truskett}\ \emph {et~al.}(2000)\citenamefont
  {Truskett}, \citenamefont {Torquato},\ and\ \citenamefont
  {Debenedetti}}]{Tr00}%
  \BibitemOpen
  \bibfield  {author} {\bibinfo {author} {\bibfnamefont {T.~M.}\ \bibnamefont
  {Truskett}}, \bibinfo {author} {\bibfnamefont {S.}~\bibnamefont {Torquato}},
  \ and\ \bibinfo {author} {\bibfnamefont {P.~G.}\ \bibnamefont
  {Debenedetti}},\ }\href@noop {} {\bibfield  {journal} {\bibinfo  {journal}
  {Phys. Rev. E}\ }\textbf {\bibinfo {volume} {62}},\ \bibinfo {pages} {993}
  (\bibinfo {year} {2000})}\BibitemShut {NoStop}%
\bibitem [{\citenamefont {Donev}\ \emph {et~al.}(2005)\citenamefont {Donev},
  \citenamefont {Torquato},\ and\ \citenamefont {Stillinger}}]{Do05}%
  \BibitemOpen
  \bibfield  {author} {\bibinfo {author} {\bibfnamefont {A.}~\bibnamefont
  {Donev}}, \bibinfo {author} {\bibfnamefont {S.}~\bibnamefont {Torquato}}, \
  and\ \bibinfo {author} {\bibfnamefont {F.~H.}\ \bibnamefont {Stillinger}},\
  }\href@noop {} {\bibfield  {journal} {\bibinfo  {journal} {Phys. Rev. E}\
  }\textbf {\bibinfo {volume} {71}},\ \bibinfo {pages} {011105} (\bibinfo
  {year} {2005})}\BibitemShut {NoStop}%
\bibitem [{\citenamefont {Hopkins}\ \emph {et~al.}(2013)\citenamefont
  {Hopkins}, \citenamefont {Stillinger},\ and\ \citenamefont
  {Torquato}}]{Ho13}%
  \BibitemOpen
  \bibfield  {author} {\bibinfo {author} {\bibfnamefont {A.~B.}\ \bibnamefont
  {Hopkins}}, \bibinfo {author} {\bibfnamefont {F.~H.}\ \bibnamefont
  {Stillinger}}, \ and\ \bibinfo {author} {\bibfnamefont {S.}~\bibnamefont
  {Torquato}},\ }\href@noop {} {\bibfield  {journal} {\bibinfo  {journal}
  {Phys. Rev. E}\ }\textbf {\bibinfo {volume} {88}},\ \bibinfo {pages} {022205}
  (\bibinfo {year} {2013})}\BibitemShut {NoStop}%
\bibitem [{\citenamefont {Jiao}\ and\ \citenamefont
  {Torquato}(2011{\natexlab{a}})}]{Ji11a}%
  \BibitemOpen
  \bibfield  {author} {\bibinfo {author} {\bibfnamefont {Y.}~\bibnamefont
  {Jiao}}\ and\ \bibinfo {author} {\bibfnamefont {S.}~\bibnamefont
  {Torquato}},\ }\href@noop {} {\bibfield  {journal} {\bibinfo  {journal}
  {Phys. Rev. E}\ }\textbf {\bibinfo {volume} {84}},\ \bibinfo {pages} {041309}
  (\bibinfo {year} {2011}{\natexlab{a}})}\BibitemShut {NoStop}%
\bibitem [{\citenamefont {Chen}\ \emph
  {et~al.}(2014{\natexlab{a}})\citenamefont {Chen}, \citenamefont {Jiao},\ and\
  \citenamefont {Torquato}}]{Ch14a}%
  \BibitemOpen
  \bibfield  {author} {\bibinfo {author} {\bibfnamefont {D.}~\bibnamefont
  {Chen}}, \bibinfo {author} {\bibfnamefont {Y.}~\bibnamefont {Jiao}}, \ and\
  \bibinfo {author} {\bibfnamefont {S.}~\bibnamefont {Torquato}},\ }\href@noop
  {} {\bibfield  {journal} {\bibinfo  {journal} {J. Phys. Chem. B}\ }\textbf
  {\bibinfo {volume} {118}},\ \bibinfo {pages} {7981} (\bibinfo {year}
  {2014}{\natexlab{a}})}\BibitemShut {NoStop}%
\bibitem [{\citenamefont {Torquato}\ \emph {et~al.}(2003)\citenamefont
  {Torquato}, \citenamefont {Donev},\ and\ \citenamefont {Stillinger}}]{To03a}%
  \BibitemOpen
  \bibfield  {author} {\bibinfo {author} {\bibfnamefont {S.}~\bibnamefont
  {Torquato}}, \bibinfo {author} {\bibfnamefont {A.}~\bibnamefont {Donev}}, \
  and\ \bibinfo {author} {\bibfnamefont {F.~H.}\ \bibnamefont {Stillinger}},\
  }\href@noop {} {\bibfield  {journal} {\bibinfo  {journal} {Int. J. Solids
  Struct.}\ }\textbf {\bibinfo {volume} {40}},\ \bibinfo {pages} {7143}
  (\bibinfo {year} {2003})}\BibitemShut {NoStop}%
\bibitem [{\citenamefont {Donev}\ \emph
  {et~al.}(2004{\natexlab{a}})\citenamefont {Donev}, \citenamefont {Torquato},
  \citenamefont {Stillinger},\ and\ \citenamefont {Connelly}}]{Do04}%
  \BibitemOpen
  \bibfield  {author} {\bibinfo {author} {\bibfnamefont {A.}~\bibnamefont
  {Donev}}, \bibinfo {author} {\bibfnamefont {S.}~\bibnamefont {Torquato}},
  \bibinfo {author} {\bibfnamefont {F.~H.}\ \bibnamefont {Stillinger}}, \ and\
  \bibinfo {author} {\bibfnamefont {R.}~\bibnamefont {Connelly}},\ }\href@noop
  {} {\bibfield  {journal} {\bibinfo  {journal} {J. Appl. Phys.}\ }\textbf
  {\bibinfo {volume} {95}},\ \bibinfo {pages} {989} (\bibinfo {year}
  {2004}{\natexlab{a}})}\BibitemShut {NoStop}%
\bibitem [{\citenamefont {Atkinson}\ \emph {et~al.}(2014)\citenamefont
  {Atkinson}, \citenamefont {Stillinger},\ and\ \citenamefont
  {Torquato}}]{At14}%
  \BibitemOpen
  \bibfield  {author} {\bibinfo {author} {\bibfnamefont {S.}~\bibnamefont
  {Atkinson}}, \bibinfo {author} {\bibfnamefont {F.~H.}\ \bibnamefont
  {Stillinger}}, \ and\ \bibinfo {author} {\bibfnamefont {S.}~\bibnamefont
  {Torquato}},\ }\href@noop {} {\bibfield  {journal} {\bibinfo  {journal}
  {Proc. Natl. Acad. Sci. USA}\ }\textbf {\bibinfo {volume} {111}},\
  \bibinfo {pages} {18436} (\bibinfo {year} {2014})}\BibitemShut {NoStop}%
\bibitem [{\citenamefont {Demchyna}\ \emph {et~al.}(2006)\citenamefont
  {Demchyna}, \citenamefont {Leoni}, \citenamefont {Rosner},\ and\
  \citenamefont {Schwarz}}]{De06}%
  \BibitemOpen
  \bibfield  {author} {\bibinfo {author} {\bibfnamefont {R.}~\bibnamefont
  {Demchyna}}, \bibinfo {author} {\bibfnamefont {S.}~\bibnamefont {Leoni}},
  \bibinfo {author} {\bibfnamefont {H.}~\bibnamefont {Rosner}}, \ and\ \bibinfo
  {author} {\bibfnamefont {U.}~\bibnamefont {Schwarz}},\ }\href@noop {}
  {\bibfield  {journal} {\bibinfo  {journal} {Z. Kristallogr.}\ }\textbf
  {\bibinfo {volume} {221}},\ \bibinfo {pages} {420} (\bibinfo {year}
  {2006})}\BibitemShut {NoStop}%
\bibitem [{\citenamefont {Cazorla}\ \emph {et~al.}(2009)\citenamefont
  {Cazorla}, \citenamefont {Errandonea},\ and\ \citenamefont {Sola}}]{Ca09}%
  \BibitemOpen
  \bibfield  {author} {\bibinfo {author} {\bibfnamefont {C.}~\bibnamefont
  {Cazorla}}, \bibinfo {author} {\bibfnamefont {D.}~\bibnamefont {Errandonea}},
  \ and\ \bibinfo {author} {\bibfnamefont {E.}~\bibnamefont {Sola}},\
  }\href@noop {} {\bibfield  {journal} {\bibinfo  {journal} {Phys. Rev. B}\
  }\textbf {\bibinfo {volume} {80}},\ \bibinfo {pages} {064105} (\bibinfo
  {year} {2009})}\BibitemShut {NoStop}%
\bibitem [{\citenamefont {Degtyareva}(2005)}]{De05}%
  \BibitemOpen
  \bibfield  {author} {\bibinfo {author} {\bibfnamefont {V.~F.}\ \bibnamefont
  {Degtyareva}},\ }\href@noop {} {\bibfield  {journal} {\bibinfo  {journal} {J.
  Synchrotron Radiat.}\ }\textbf {\bibinfo {volume} {12}},\ \bibinfo {pages}
  {584} (\bibinfo {year} {2005})}\BibitemShut {NoStop}%
\bibitem [{\citenamefont {Kochevets}\ \emph {et~al.}(2001)\citenamefont
  {Kochevets}, \citenamefont {Buckmaster}, \citenamefont {Jackson},\ and\
  \citenamefont {Hegab}}]{Ko01}%
  \BibitemOpen
  \bibfield  {author} {\bibinfo {author} {\bibfnamefont {S.}~\bibnamefont
  {Kochevets}}, \bibinfo {author} {\bibfnamefont {J.}~\bibnamefont
  {Buckmaster}}, \bibinfo {author} {\bibfnamefont {T.~L.}\ \bibnamefont
  {Jackson}}, \ and\ \bibinfo {author} {\bibfnamefont {A.}~\bibnamefont
  {Hegab}},\ }\href@noop {} {\bibfield  {journal} {\bibinfo  {journal} {J.
  Propul. Power}\ }\textbf {\bibinfo {volume} {17}},\ \bibinfo {pages} {883}
  (\bibinfo {year} {2001})}\BibitemShut {NoStop}%
\bibitem [{\citenamefont {Mindess}\ and\ \citenamefont {Young}(1981)}]{Mi81}%
  \BibitemOpen
  \bibfield  {author} {\bibinfo {author} {\bibfnamefont {S.}~\bibnamefont
  {Mindess}}\ and\ \bibinfo {author} {\bibfnamefont {J.~F.}\ \bibnamefont
  {Young}},\ }\href@noop {} {\emph {\bibinfo {title} {Concrete}}}\ (\bibinfo
  {publisher} {Prentice-Hall, Englewood Cliffs, NJ},\ \bibinfo {year}
  {1981})\BibitemShut {NoStop}%
\bibitem [{\citenamefont {Kingery}\ \emph {et~al.}(1976)\citenamefont
  {Kingery}, \citenamefont {Bowen},\ and\ \citenamefont {Uhlmann}}]{Ki76}%
  \BibitemOpen
  \bibfield  {author} {\bibinfo {author} {\bibfnamefont {W.~D.}\ \bibnamefont
  {Kingery}}, \bibinfo {author} {\bibfnamefont {H.~K.}\ \bibnamefont {Bowen}},
  \ and\ \bibinfo {author} {\bibfnamefont {D.~R.}\ \bibnamefont {Uhlmann}},\
  }\href@noop {} {\emph {\bibinfo {title} {Introduction to Ceramics}}},\
  \bibinfo {edition} {2nd}\ ed.\ (\bibinfo  {publisher} {Wiley-Interscience,
  New York},\ \bibinfo {year} {1976})\BibitemShut {NoStop}%
\bibitem [{\citenamefont {Spangenberg}\ \emph {et~al.}(2014)\citenamefont
  {Spangenberg}, \citenamefont {Scherer}, \citenamefont {Hopkins},\ and\
  \citenamefont {Torquato}}]{Sp14}%
  \BibitemOpen
  \bibfield  {author} {\bibinfo {author} {\bibfnamefont {J.}~\bibnamefont
  {Spangenberg}}, \bibinfo {author} {\bibfnamefont {G.~W.}\ \bibnamefont
  {Scherer}}, \bibinfo {author} {\bibfnamefont {A.~B.}\ \bibnamefont
  {Hopkins}}, \ and\ \bibinfo {author} {\bibfnamefont {S.}~\bibnamefont
  {Torquato}},\ }\href@noop {} {\bibfield  {journal} {\bibinfo  {journal} {J.
  Appl. Phys.}\ }\textbf {\bibinfo {volume} {116}},\ \bibinfo {pages} {184902}
  (\bibinfo {year} {2014})}\BibitemShut {NoStop}%
\bibitem [{\citenamefont {Larcher}\ and\ \citenamefont {Jenkins}(2015)}]{La15}%
  \BibitemOpen
  \bibfield  {author} {\bibinfo {author} {\bibfnamefont {M.}~\bibnamefont
  {Larcher}}\ and\ \bibinfo {author} {\bibfnamefont {J.~T.}\ \bibnamefont
  {Jenkins}},\ }\href@noop {} {\bibfield  {journal} {\bibinfo  {journal} {J.
  Fluid Mech.}\ }\textbf {\bibinfo {volume} {782}},\ \bibinfo {pages} {405}
  (\bibinfo {year} {2015})}\BibitemShut {NoStop}%
\bibitem [{\citenamefont {Hopkins}\ \emph {et~al.}(2010)\citenamefont
  {Hopkins}, \citenamefont {Stillinger},\ and\ \citenamefont
  {Torquato}}]{Ho10}%
  \BibitemOpen
  \bibfield  {author} {\bibinfo {author} {\bibfnamefont {A.~B.}\ \bibnamefont
  {Hopkins}}, \bibinfo {author} {\bibfnamefont {F.~H.}\ \bibnamefont
  {Stillinger}}, \ and\ \bibinfo {author} {\bibfnamefont {S.}~\bibnamefont
  {Torquato}},\ }\href@noop {} {\bibfield  {journal} {\bibinfo  {journal}
  {Phys. Rev. E}\ }\textbf {\bibinfo {volume} {81}},\ \bibinfo {pages} {041305}
  (\bibinfo {year} {2010})}\BibitemShut {NoStop}%
\bibitem [{\citenamefont {Hopkins}\ \emph {et~al.}(2011)\citenamefont
  {Hopkins}, \citenamefont {Stillinger},\ and\ \citenamefont
  {Torquato}}]{Ho11}%
  \BibitemOpen
  \bibfield  {author} {\bibinfo {author} {\bibfnamefont {A.~B.}\ \bibnamefont
  {Hopkins}}, \bibinfo {author} {\bibfnamefont {F.~H.}\ \bibnamefont
  {Stillinger}}, \ and\ \bibinfo {author} {\bibfnamefont {S.}~\bibnamefont
  {Torquato}},\ }\href@noop {} {\bibfield  {journal} {\bibinfo  {journal}
  {Phys. Rev. E}\ }\textbf {\bibinfo {volume} {83}},\ \bibinfo {pages} {011304}
  (\bibinfo {year} {2011})}\BibitemShut {NoStop}%
\bibitem [{\citenamefont {Courtemanche}\ \emph {et~al.}(1993)\citenamefont
  {Courtemanche}, \citenamefont {Pasmore},\ and\ \citenamefont
  {Van~Swol}}]{Co93}%
  \BibitemOpen
  \bibfield  {author} {\bibinfo {author} {\bibfnamefont {D.~J.}\ \bibnamefont
  {Courtemanche}}, \bibinfo {author} {\bibfnamefont {T.~A.}\ \bibnamefont
  {Pasmore}}, \ and\ \bibinfo {author} {\bibfnamefont {F.}~\bibnamefont
  {Van~Swol}},\ }\href@noop {} {\bibfield  {journal} {\bibinfo  {journal} {Mol.
  Phys.}\ }\textbf {\bibinfo {volume} {80}},\ \bibinfo {pages} {861} (\bibinfo
  {year} {1993})}\BibitemShut {NoStop}%
\bibitem [{\citenamefont {Schmidt}\ and\ \citenamefont
  {L{\"o}wen}(1997)}]{Sc97}%
  \BibitemOpen
  \bibfield  {author} {\bibinfo {author} {\bibfnamefont {M.}~\bibnamefont
  {Schmidt}}\ and\ \bibinfo {author} {\bibfnamefont {H.}~\bibnamefont
  {L{\"o}wen}},\ }\href@noop {} {\bibfield  {journal} {\bibinfo  {journal}
  {Phys. Rev. E}\ }\textbf {\bibinfo {volume} {55}},\ \bibinfo {pages} {7228}
  (\bibinfo {year} {1997})}\BibitemShut {NoStop}%
\bibitem{Fo06}
A. Fortini and M. Dijkstra, J. Phys.: Condens. Matter {\bf 18}, L371 (2006); W. Qi, Y. Peng, Y. Han, R. K. Bowles, and M. Dijkstra, Phys. Rev. Lett. {\bf 115}, 185701 (2015).
\bibitem [{\citenamefont {O{\u{g}}uz}\ \emph {et~al.}(2012)\citenamefont
  {O{\u{g}}uz}, \citenamefont {Marechal}, \citenamefont {Ramiro-Manzano},
  \citenamefont {Rodriguez}, \citenamefont {Messina}, \citenamefont
  {Meseguer},\ and\ \citenamefont {L{\"o}wen}}]{Og12}%
  \BibitemOpen
  \bibfield  {author} {\bibinfo {author} {\bibfnamefont {E.~C.}\ \bibnamefont
  {O{\u{g}}uz}}, \bibinfo {author} {\bibfnamefont {M.}~\bibnamefont
  {Marechal}}, \bibinfo {author} {\bibfnamefont {F.}~\bibnamefont
  {Ramiro-Manzano}}, \bibinfo {author} {\bibfnamefont {I.}~\bibnamefont
  {Rodriguez}}, \bibinfo {author} {\bibfnamefont {R.}~\bibnamefont {Messina}},
  \bibinfo {author} {\bibfnamefont {F.~J.}\ \bibnamefont {Meseguer}}, \ and\
  \bibinfo {author} {\bibfnamefont {H.}~\bibnamefont {L{\"o}wen}},\ }\href@noop
  {} {\bibfield  {journal} {\bibinfo  {journal} {Phys. Rev. Lett.}\ }\textbf
  {\bibinfo {volume} {109}},\ \bibinfo {pages} {218301} (\bibinfo {year}
  {2012})}\BibitemShut {NoStop}%
\bibitem [{\citenamefont {Yamchi}\ and\ \citenamefont {Bowles}(2015)}]{Ya15}%
  \BibitemOpen
  \bibfield  {author} {\bibinfo {author} {\bibfnamefont {M.~Z.}\ \bibnamefont
  {Yamchi}}\ and\ \bibinfo {author} {\bibfnamefont {R.~K.}\ \bibnamefont
  {Bowles}},\ }\href@noop {} {\bibfield  {journal} {\bibinfo  {journal} {Phys.
  Rev. Lett.}\ }\textbf {\bibinfo {volume} {115}},\ \bibinfo {pages} {025702}
  (\bibinfo {year} {2015})}\BibitemShut {NoStop}%
\bibitem [{\citenamefont {Abdeljawad}\ \emph {et~al.}(2014)\citenamefont
  {Abdeljawad}, \citenamefont {V{\"o}lker}, \citenamefont {Davis},
  \citenamefont {McMeeking},\ and\ \citenamefont {Haataja}}]{Ab14}%
  \BibitemOpen
  \bibfield  {author} {\bibinfo {author} {\bibfnamefont {F.}~\bibnamefont
  {Abdeljawad}}, \bibinfo {author} {\bibfnamefont {B.}~\bibnamefont
  {V{\"o}lker}}, \bibinfo {author} {\bibfnamefont {R.}~\bibnamefont {Davis}},
  \bibinfo {author} {\bibfnamefont {R.~M.}\ \bibnamefont {McMeeking}}, \ and\
  \bibinfo {author} {\bibfnamefont {M.}~\bibnamefont {Haataja}},\ }\href@noop
  {} {\bibfield  {journal} {\bibinfo  {journal} {J. Power Sources}\ }\textbf
  {\bibinfo {volume} {250}},\ \bibinfo {pages} {319} (\bibinfo {year}
  {2014})}\BibitemShut {NoStop}%
\bibitem [{\citenamefont {Minton}(1992)}]{Mi92}%
  \BibitemOpen
  \bibfield  {author} {\bibinfo {author} {\bibfnamefont {A.~P.}\ \bibnamefont
  {Minton}},\ }\href@noop {} {\bibfield  {journal} {\bibinfo  {journal}
  {Biophys. J.}\ }\textbf {\bibinfo {volume} {63}},\ \bibinfo {pages} {1090}
  (\bibinfo {year} {1992})}\BibitemShut {NoStop}%
\bibitem [{\citenamefont {Ellis}(2001)}]{El01}%
  \BibitemOpen
  \bibfield  {author} {\bibinfo {author} {\bibfnamefont {R.~J.}\ \bibnamefont
  {Ellis}},\ }\href@noop {} {\bibfield  {journal} {\bibinfo  {journal} {Trends
  Biochem. Sci.}\ }\textbf {\bibinfo {volume} {26}},\ \bibinfo {pages} {597}
  (\bibinfo {year} {2001})}\BibitemShut {NoStop}%
\bibitem [{\citenamefont {Gevertz}\ and\ \citenamefont
  {Torquato}(2008)}]{Ge08}%
  \BibitemOpen
  \bibfield  {author} {\bibinfo {author} {\bibfnamefont {J.~L.}\ \bibnamefont
  {Gevertz}}\ and\ \bibinfo {author} {\bibfnamefont {S.}~\bibnamefont
  {Torquato}},\ }\href@noop {} {\bibfield  {journal} {\bibinfo  {journal} {PLoS
  Comput. Biol.}\ }\textbf {\bibinfo {volume} {4}},\ \bibinfo {pages}
  {e1000152} (\bibinfo {year} {2008})}\BibitemShut {NoStop}%
\bibitem [{\citenamefont {Jiao}\ \emph {et~al.}(2014)\citenamefont {Jiao},
  \citenamefont {Lau}, \citenamefont {Hatzikirou}, \citenamefont
  {Meyer-Hermann}, \citenamefont {Corbo},\ and\ \citenamefont
  {Torquato}}]{Ji14}%
  \BibitemOpen
  \bibfield  {author} {\bibinfo {author} {\bibfnamefont {Y.}~\bibnamefont
  {Jiao}}, \bibinfo {author} {\bibfnamefont {T.}~\bibnamefont {Lau}}, \bibinfo
  {author} {\bibfnamefont {H.}~\bibnamefont {Hatzikirou}}, \bibinfo {author}
  {\bibfnamefont {M.}~\bibnamefont {Meyer-Hermann}}, \bibinfo {author}
  {\bibfnamefont {J.~C.}\ \bibnamefont {Corbo}}, \ and\ \bibinfo {author}
  {\bibfnamefont {S.}~\bibnamefont {Torquato}},\ }\href@noop {} {\bibfield
  {journal} {\bibinfo  {journal} {Phys. Rev. E}\ }\textbf {\bibinfo {volume}
  {89}},\ \bibinfo {pages} {022721} (\bibinfo {year} {2014})}\BibitemShut
  {NoStop}%
\bibitem [{\citenamefont {Popov}\ \emph {et~al.}(2014)\citenamefont {Popov},
  \citenamefont {Kovalski}, \citenamefont {Grandi}, \citenamefont {Bagnoli},\
  and\ \citenamefont {Amieva}}]{Po14}%
  \BibitemOpen
  \bibfield  {author} {\bibinfo {author} {\bibfnamefont {L.}~\bibnamefont
  {Popov}}, \bibinfo {author} {\bibfnamefont {J.}~\bibnamefont {Kovalski}},
  \bibinfo {author} {\bibfnamefont {G.}~\bibnamefont {Grandi}}, \bibinfo
  {author} {\bibfnamefont {F.}~\bibnamefont {Bagnoli}}, \ and\ \bibinfo
  {author} {\bibfnamefont {M.~R.}\ \bibnamefont {Amieva}},\ }\href@noop {}
  {\bibfield  {journal} {\bibinfo  {journal} {Front. Immunol.}\ }\textbf
  {\bibinfo {volume} {5}},\ \bibinfo {pages} {41} (\bibinfo {year}
  {2014})}\BibitemShut {NoStop}%
\bibitem{Pr04}
R. O. Prum, J. A. Cole, and R. H. Torres, J. Exp. Biol. {\bf 207}, 3999 (2004); B. Q. Dong, X. H. Liu, T. R. Zhan, L. P. Jiang, H. W. Yin, F. Liu, and J. Zi, Opt. Express {\bf 18}, 14430 (2010).
\bibitem [{\citenamefont {Torquato}(2011)}]{To11}%
  \BibitemOpen
  \bibfield  {author} {\bibinfo {author} {\bibfnamefont {S.}~\bibnamefont
  {Torquato}},\ }\href@noop {} {\bibfield  {journal} {\bibinfo  {journal}
  {Phys. Biol.}\ }\textbf {\bibinfo {volume} {8}},\ \bibinfo {pages} {015017}
  (\bibinfo {year} {2011})}\BibitemShut {NoStop}%
\bibitem [{\citenamefont {Jiao}\ and\ \citenamefont
  {Torquato}(2011{\natexlab{b}})}]{Ji11b}%
  \BibitemOpen
  \bibfield  {author} {\bibinfo {author} {\bibfnamefont {Y.}~\bibnamefont
  {Jiao}}\ and\ \bibinfo {author} {\bibfnamefont {S.}~\bibnamefont
  {Torquato}},\ }\href@noop {} {\bibfield  {journal} {\bibinfo  {journal} {PLoS
  Comput. Biol.}\ }\textbf {\bibinfo {volume} {7}},\ \bibinfo {pages}
  {e1002314} (\bibinfo {year} {2011}{\natexlab{b}})}\BibitemShut {NoStop}%
\bibitem [{\citenamefont {Jiao}\ and\ \citenamefont {Torquato}(2013)}]{Ji13}%
  \BibitemOpen
  \bibfield  {author} {\bibinfo {author} {\bibfnamefont {Y.}~\bibnamefont
  {Jiao}}\ and\ \bibinfo {author} {\bibfnamefont {S.}~\bibnamefont
  {Torquato}},\ }\href@noop {} {\bibfield  {journal} {\bibinfo  {journal}
  {Phys. Rev. E}\ }\textbf {\bibinfo {volume} {87}},\ \bibinfo {pages} {052707}
  (\bibinfo {year} {2013})}\BibitemShut {NoStop}%
\bibitem [{\citenamefont {Chen}\ \emph
  {et~al.}(2014{\natexlab{b}})\citenamefont {Chen}, \citenamefont {Jiao},\ and\
  \citenamefont {Torquato}}]{Ch14b}%
  \BibitemOpen
  \bibfield  {author} {\bibinfo {author} {\bibfnamefont {D.}~\bibnamefont
  {Chen}}, \bibinfo {author} {\bibfnamefont {Y.}~\bibnamefont {Jiao}}, \ and\
  \bibinfo {author} {\bibfnamefont {S.}~\bibnamefont {Torquato}},\ }\href@noop
  {} {\bibfield  {journal} {\bibinfo  {journal} {PLoS ONE}\ }\textbf {\bibinfo
  {volume} {9}},\ \bibinfo {pages} {e109934} (\bibinfo {year}
  {2014}{\natexlab{b}})}\BibitemShut {NoStop}%
\bibitem [{\citenamefont {Landry}\ \emph {et~al.}(2004)\citenamefont {Landry},
  \citenamefont {Grest},\ and\ \citenamefont {Plimpton}}]{La04}%
  \BibitemOpen
  \bibfield  {author} {\bibinfo {author} {\bibfnamefont {J.~W.}\ \bibnamefont
  {Landry}}, \bibinfo {author} {\bibfnamefont {G.~S.}\ \bibnamefont {Grest}}, \
  and\ \bibinfo {author} {\bibfnamefont {S.~J.}\ \bibnamefont {Plimpton}},\
  }\href@noop {} {\bibfield  {journal} {\bibinfo  {journal} {Powder Technol.}\
  }\textbf {\bibinfo {volume} {139}},\ \bibinfo {pages} {233} (\bibinfo {year}
  {2004})}\BibitemShut {NoStop}%
\bibitem [{\citenamefont {Desmond}\ and\ \citenamefont {Weeks}(2009)}]{De09}%
  \BibitemOpen
  \bibfield  {author} {\bibinfo {author} {\bibfnamefont {K.~W.}\ \bibnamefont
  {Desmond}}\ and\ \bibinfo {author} {\bibfnamefont {E.~R.}\ \bibnamefont
  {Weeks}},\ }\href@noop {} {\bibfield  {journal} {\bibinfo  {journal} {Phys.
  Rev. E}\ }\textbf {\bibinfo {volume} {80}},\ \bibinfo {pages} {051305}
  (\bibinfo {year} {2009})}\BibitemShut {NoStop}%
\bibitem [{\citenamefont {Torquato}\ and\ \citenamefont
  {Stillinger}(2001)}]{To01}%
  \BibitemOpen
  \bibfield  {author} {\bibinfo {author} {\bibfnamefont {S.}~\bibnamefont
  {Torquato}}\ and\ \bibinfo {author} {\bibfnamefont {F.~H.}\ \bibnamefont
  {Stillinger}},\ }\href@noop {} {\bibfield  {journal} {\bibinfo  {journal} {J.
  Phys. Chem. B}\ }\textbf {\bibinfo {volume} {105}},\ \bibinfo {pages} {11849}
  (\bibinfo {year} {2001})}\BibitemShut {NoStop}%
\bibitem [{Note1()}]{Note1}%
  \BibitemOpen
  \bibinfo {note} {We refer to these generated packings as MRJ due to the
  ability of the TJ algorithm and the use of RSA initial conditions to produce
  them with high fidelity in the bulk \cite {At13, Ho13}. A detailed study
  verifying that these are indeed MRJ states by characterizing their structural
  disorder via a variety of order metrics is deferred to a future work due to
  the challenging task of enumerating jammed states for each combination of
  $H$, $\alpha $ and $x$.}\BibitemShut {Stop}%
\bibitem [{\citenamefont {Hopkins}\ \emph {et~al.}(2012)\citenamefont
  {Hopkins}, \citenamefont {Stillinger},\ and\ \citenamefont
  {Torquato}}]{Ho12}%
  \BibitemOpen
  \bibfield  {author} {\bibinfo {author} {\bibfnamefont {A.~B.}\ \bibnamefont
  {Hopkins}}, \bibinfo {author} {\bibfnamefont {F.~H.}\ \bibnamefont
  {Stillinger}}, \ and\ \bibinfo {author} {\bibfnamefont {S.}~\bibnamefont
  {Torquato}},\ }\href@noop {} {\bibfield  {journal} {\bibinfo  {journal}
  {Phys. Rev. E}\ }\textbf {\bibinfo {volume} {85}},\ \bibinfo {pages} {021130}
  (\bibinfo {year} {2012})}\BibitemShut {NoStop}%
\bibitem{To03b}
S. Torquato and F. H. Stillinger, Phys. Rev. E, {\bf 68}, 041113 (2003).
\bibitem [{\citenamefont {Dufresne}\ \emph {et~al.}(2009)\citenamefont
  {Dufresne}, \citenamefont {Noh}, \citenamefont {Saranathan}, \citenamefont
  {Mochrie}, \citenamefont {Cao},\ and\ \citenamefont {Prum}}]{Du09}%
  \BibitemOpen
  \bibfield  {author} {\bibinfo {author} {\bibfnamefont {E.~R.}\ \bibnamefont
  {Dufresne}}, \bibinfo {author} {\bibfnamefont {H.}~\bibnamefont {Noh}},
  \bibinfo {author} {\bibfnamefont {V.}~\bibnamefont {Saranathan}}, \bibinfo
  {author} {\bibfnamefont {S.~G.~J.}\ \bibnamefont {Mochrie}}, \bibinfo
  {author} {\bibfnamefont {H.}~\bibnamefont {Cao}}, \ and\ \bibinfo {author}
  {\bibfnamefont {R.~O.}\ \bibnamefont {Prum}},\ }\href@noop {} {\bibfield
  {journal} {\bibinfo  {journal} {Soft Matter}\ }\textbf {\bibinfo {volume}
  {5}},\ \bibinfo {pages} {1792} (\bibinfo {year} {2009})}\BibitemShut
  {NoStop}%
\bibitem [{\citenamefont {Torquato}\ and\ \citenamefont
  {Jiao}(2009{\natexlab{a}})}]{To09a}%
  \BibitemOpen
  \bibfield  {author} {\bibinfo {author} {\bibfnamefont {S.}~\bibnamefont
  {Torquato}}\ and\ \bibinfo {author} {\bibfnamefont {Y.}~\bibnamefont
  {Jiao}},\ }\href@noop {} {\bibfield  {journal} {\bibinfo  {journal} {Nature (London)}\
  }\textbf {\bibinfo {volume} {460}},\ \bibinfo {pages} {876} (\bibinfo {year}
  {2009}{\natexlab{a}})}\BibitemShut {NoStop}%
\bibitem [{\citenamefont {Torquato}\ and\ \citenamefont
  {Jiao}(2009{\natexlab{b}})}]{To09b}%
  \BibitemOpen
  \bibfield  {author} {\bibinfo {author} {\bibfnamefont {S.}~\bibnamefont
  {Torquato}}\ and\ \bibinfo {author} {\bibfnamefont {Y.}~\bibnamefont
  {Jiao}},\ }\href@noop {} {\bibfield  {journal} {\bibinfo  {journal} {Phys.
  Rev. E}\ }\textbf {\bibinfo {volume} {80}},\ \bibinfo {pages} {041104}
  (\bibinfo {year} {2009}{\natexlab{b}})}\BibitemShut {NoStop}%
\bibitem [{\citenamefont {Marcotte}\ and\ \citenamefont
  {Torquato}(2013)}]{Ma13}%
  \BibitemOpen
  \bibfield  {author} {\bibinfo {author} {\bibfnamefont {{\'E}.}~\bibnamefont
  {Marcotte}}\ and\ \bibinfo {author} {\bibfnamefont {S.}~\bibnamefont
  {Torquato}},\ }\href@noop {} {\bibfield  {journal} {\bibinfo  {journal}
  {Phys. Rev. E}\ }\textbf {\bibinfo {volume} {87}},\ \bibinfo {pages} {063303}
  (\bibinfo {year} {2013})}\BibitemShut {NoStop}%
\bibitem [{\citenamefont {Lubachevsky}\ \emph {et~al.}(1991)\citenamefont
  {Lubachevsky}, \citenamefont {Stillinger},\ and\ \citenamefont
  {Pinson}}]{Lu91}%
  \BibitemOpen
  \bibfield  {author} {\bibinfo {author} {\bibfnamefont {B.~D.}\ \bibnamefont
  {Lubachevsky}}, \bibinfo {author} {\bibfnamefont {F.~H.}\ \bibnamefont
  {Stillinger}}, \ and\ \bibinfo {author} {\bibfnamefont {E.~N.}\ \bibnamefont
  {Pinson}},\ }\href@noop {} {\bibfield  {journal} {\bibinfo  {journal} {J.
  Stat. Phys.}\ }\textbf {\bibinfo {volume} {64}},\ \bibinfo {pages} {501}
  (\bibinfo {year} {1991})}\BibitemShut {NoStop}%
\bibitem{Bi09}
I. Biazzo, F. Caltagirone, G. Parisi, and F. Zamponi, Phys. Rev. Lett., {\bf 102}, 195701 (2009).
\bibitem{Da10}
M. Danisch, Y. Jin, and H. A. Makse, Phys. Rev. E, {\bf 81}, 051303 (2010).
\bibitem [{\citenamefont {Lu}\ and\ \citenamefont {Torquato}(1990)}]{Lu90}%
  \BibitemOpen
  \bibfield  {author} {\bibinfo {author} {\bibfnamefont {B.}~\bibnamefont
  {Lu}}\ and\ \bibinfo {author} {\bibfnamefont {S.}~\bibnamefont {Torquato}},\
  }\href@noop {} {\bibfield  {journal} {\bibinfo  {journal} {J. Chem. Phys.}\
  }\textbf {\bibinfo {volume} {93}},\ \bibinfo {pages} {3452} (\bibinfo {year}
  {1990})}\BibitemShut {NoStop}%
\bibitem [{\citenamefont {Zachary}\ and\ \citenamefont
  {Torquato}(2009)}]{Za09}%
  \BibitemOpen
  \bibfield  {author} {\bibinfo {author} {\bibfnamefont {C.~E.}\ \bibnamefont
  {Zachary}}\ and\ \bibinfo {author} {\bibfnamefont {S.}~\bibnamefont
  {Torquato}},\ }\href@noop {} {\bibfield  {journal} {\bibinfo  {journal} {J.
  Stat. Mech. Theor. Exp.}\ }\textbf {\bibinfo {volume} {2009}},\ \bibinfo
  {pages} {P12015} (\bibinfo {year} {2009})}\BibitemShut {NoStop}%
\bibitem [{\citenamefont {Zachary}\ \emph
  {et~al.}(2011{\natexlab{a}})\citenamefont {Zachary}, \citenamefont {Jiao},\
  and\ \citenamefont {Torquato}}]{Za11a}%
  \BibitemOpen
  \bibfield  {author} {\bibinfo {author} {\bibfnamefont {C.~E.}\ \bibnamefont
  {Zachary}}, \bibinfo {author} {\bibfnamefont {Y.}~\bibnamefont {Jiao}}, \
  and\ \bibinfo {author} {\bibfnamefont {S.}~\bibnamefont {Torquato}},\
  }\href@noop {} {\bibfield  {journal} {\bibinfo  {journal} {Phys. Rev. Lett.}\
  }\textbf {\bibinfo {volume} {106}},\ \bibinfo {pages} {178001} (\bibinfo
  {year} {2011}{\natexlab{a}})}\BibitemShut {NoStop}%
\bibitem [{\citenamefont {Zachary}\ \emph
  {et~al.}(2011{\natexlab{b}})\citenamefont {Zachary}, \citenamefont {Jiao},\
  and\ \citenamefont {Torquato}}]{Za11b}%
  \BibitemOpen
  \bibfield  {author} {\bibinfo {author} {\bibfnamefont {C.~E.}\ \bibnamefont
  {Zachary}}, \bibinfo {author} {\bibfnamefont {Y.}~\bibnamefont {Jiao}}, \
  and\ \bibinfo {author} {\bibfnamefont {S.}~\bibnamefont {Torquato}},\
  }\href@noop {} {\bibfield  {journal} {\bibinfo  {journal} {Phys. Rev. E}\
  }\textbf {\bibinfo {volume} {83}},\ \bibinfo {pages} {051308} (\bibinfo
  {year} {2011}{\natexlab{b}})}\BibitemShut {NoStop}%
\bibitem [{\citenamefont {Zachary}\ \emph
  {et~al.}(2011{\natexlab{c}})\citenamefont {Zachary}, \citenamefont {Jiao},\
  and\ \citenamefont {Torquato}}]{Za11c}%
  \BibitemOpen
  \bibfield  {author} {\bibinfo {author} {\bibfnamefont {C.~E.}\ \bibnamefont
  {Zachary}}, \bibinfo {author} {\bibfnamefont {Y.}~\bibnamefont {Jiao}}, \
  and\ \bibinfo {author} {\bibfnamefont {S.}~\bibnamefont {Torquato}},\
  }\href@noop {} {\bibfield  {journal} {\bibinfo  {journal} {Phys. Rev. E}\
  }\textbf {\bibinfo {volume} {83}},\ \bibinfo {pages} {051309} (\bibinfo
  {year} {2011}{\natexlab{c}})}\BibitemShut {NoStop}%
\bibitem [{\citenamefont {Dreyfus}\ \emph {et~al.}(2015)\citenamefont
  {Dreyfus}, \citenamefont {Xu}, \citenamefont {Still}, \citenamefont {Hough},
  \citenamefont {Yodh},\ and\ \citenamefont {Torquato}}]{Dr15}%
  \BibitemOpen
  \bibfield  {author} {\bibinfo {author} {\bibfnamefont {R.}~\bibnamefont
  {Dreyfus}}, \bibinfo {author} {\bibfnamefont {Y.}~\bibnamefont {Xu}},
  \bibinfo {author} {\bibfnamefont {T.}~\bibnamefont {Still}}, \bibinfo
  {author} {\bibfnamefont {L.~A.}\ \bibnamefont {Hough}}, \bibinfo {author}
  {\bibfnamefont {A.~G.}\ \bibnamefont {Yodh}}, \ and\ \bibinfo {author}
  {\bibfnamefont {S.}~\bibnamefont {Torquato}},\ }\href@noop {} {\bibfield
  {journal} {\bibinfo  {journal} {Phys. Rev. E}\ }\textbf {\bibinfo {volume}
  {91}},\ \bibinfo {pages} {012302} (\bibinfo {year} {2015})}\BibitemShut
  {NoStop}%
\bibitem [{Note2()}]{Note2}%
  \BibitemOpen
\bibfield  {journal} {  }\bibinfo {note} {This is not surprising since $\sigma
  _{\tau }^2(R)$ effectively homogenizes information at different heights
  between the two planes, which leads to its insensitivity to
  confinement.}\BibitemShut {Stop}%
\bibitem [{\citenamefont {Shibuta}(2012)}]{Sh12}%
  \BibitemOpen
  \bibfield  {author} {\bibinfo {author} {\bibfnamefont {Y.}~\bibnamefont
  {Shibuta}},\ }\href@noop {} {\bibfield  {journal} {\bibinfo  {journal} {Chem.
  Phys. Lett.}\ }\textbf {\bibinfo {volume} {532}},\ \bibinfo {pages} {84}
  (\bibinfo {year} {2012})}\BibitemShut {NoStop}%
\bibitem [{\citenamefont {Foteinopoulou}\ \emph {et~al.}(2015)\citenamefont
  {Foteinopoulou}, \citenamefont {Karayiannis},\ and\ \citenamefont
  {Laso}}]{Fo15}%
  \BibitemOpen
  \bibfield  {author} {\bibinfo {author} {\bibfnamefont {K.}~\bibnamefont
  {Foteinopoulou}}, \bibinfo {author} {\bibfnamefont {N.~C.}\ \bibnamefont
  {Karayiannis}}, \ and\ \bibinfo {author} {\bibfnamefont {M.}~\bibnamefont
  {Laso}},\ }\href@noop {} {\bibfield  {journal} {\bibinfo  {journal} {Chem.
  Eng. Sci.}\ }\textbf {\bibinfo {volume} {121}},\ \bibinfo {pages} {118}
  (\bibinfo {year} {2015})}\BibitemShut {NoStop}%
\bibitem [{\citenamefont {Mirny}(2011)}]{Mi11}%
  \BibitemOpen
  \bibfield  {author} {\bibinfo {author} {\bibfnamefont {L.~A.}\ \bibnamefont
  {Mirny}},\ }\href@noop {} {\bibfield  {journal} {\bibinfo  {journal}
  {Chromosome Res.}\ }\textbf {\bibinfo {volume} {19}},\ \bibinfo {pages} {37}
  (\bibinfo {year} {2011})}\BibitemShut {NoStop}%
\bibitem [{\citenamefont {Donev}\ \emph
  {et~al.}(2004{\natexlab{b}})\citenamefont {Donev}, \citenamefont {Cisse},
  \citenamefont {Sachs}, \citenamefont {Variano}, \citenamefont {Stillinger},
  \citenamefont {Connelly}, \citenamefont {Torquato},\ and\ \citenamefont
  {Chaikin}}]{Do04b}%
  \BibitemOpen
  \bibfield  {author} {\bibinfo {author} {\bibfnamefont {A.}~\bibnamefont
  {Donev}}, \bibinfo {author} {\bibfnamefont {I.}~\bibnamefont {Cisse}},
  \bibinfo {author} {\bibfnamefont {D.}~\bibnamefont {Sachs}}, \bibinfo
  {author} {\bibfnamefont {E.~A.}\ \bibnamefont {Variano}}, \bibinfo {author}
  {\bibfnamefont {F.~H.}\ \bibnamefont {Stillinger}}, \bibinfo {author}
  {\bibfnamefont {R.}~\bibnamefont {Connelly}}, \bibinfo {author}
  {\bibfnamefont {S.}~\bibnamefont {Torquato}}, \ and\ \bibinfo {author}
  {\bibfnamefont {P.~M.}\ \bibnamefont {Chaikin}},\ }\href@noop {} {\bibfield
  {journal} {\bibinfo  {journal} {Science}\ }\textbf {\bibinfo {volume}
  {303}},\ \bibinfo {pages} {990} (\bibinfo {year}
  {2004}{\natexlab{b}})}\BibitemShut {NoStop}%
\bibitem [{\citenamefont {Man}\ \emph {et~al.}(2005)\citenamefont {Man},
  \citenamefont {Donev}, \citenamefont {Stillinger}, \citenamefont {Sullivan},
  \citenamefont {Russel}, \citenamefont {Heeger}, \citenamefont {Inati},
  \citenamefont {Torquato},\ and\ \citenamefont {Chaikin}}]{Ma05}%
  \BibitemOpen
  \bibfield  {author} {\bibinfo {author} {\bibfnamefont {W.}~\bibnamefont
  {Man}}, \bibinfo {author} {\bibfnamefont {A.}~\bibnamefont {Donev}}, \bibinfo
  {author} {\bibfnamefont {F.~H.}\ \bibnamefont {Stillinger}}, \bibinfo
  {author} {\bibfnamefont {M.~T.}\ \bibnamefont {Sullivan}}, \bibinfo {author}
  {\bibfnamefont {W.~B.}\ \bibnamefont {Russel}}, \bibinfo {author}
  {\bibfnamefont {D.}~\bibnamefont {Heeger}}, \bibinfo {author} {\bibfnamefont
  {S.}~\bibnamefont {Inati}}, \bibinfo {author} {\bibfnamefont
  {S.}~\bibnamefont {Torquato}}, \ and\ \bibinfo {author} {\bibfnamefont
  {P.~M.}\ \bibnamefont {Chaikin}},\ }\href@noop {} {\bibfield  {journal}
  {\bibinfo  {journal} {Phys. Rev. Lett.}\ }\textbf {\bibinfo {volume} {94}},\
  \bibinfo {pages} {198001} (\bibinfo {year} {2005})}\BibitemShut {NoStop}%
\bibitem [{\citenamefont {Schaller}\ \emph {et~al.}(2015)\citenamefont
  {Schaller}, \citenamefont {Kapfer}, \citenamefont {Hilton}, \citenamefont
  {Cleary}, \citenamefont {Mecke}, \citenamefont {De~Michele}, \citenamefont
  {Schilling}, \citenamefont {Saadatfar}, \citenamefont {Schr{\"o}ter},
  \citenamefont {Delaney} \emph {et~al.}}]{Sc15}%
  \BibitemOpen
  \bibfield  {author} {\bibinfo {author} {\bibfnamefont {F.~M.}\ \bibnamefont
  {Schaller}}, \bibinfo {author} {\bibfnamefont {S.~C.}\ \bibnamefont
  {Kapfer}}, \bibinfo {author} {\bibfnamefont {J.~E.}\ \bibnamefont {Hilton}},
  \bibinfo {author} {\bibfnamefont {P.~W.}\ \bibnamefont {Cleary}}, \bibinfo
  {author} {\bibfnamefont {K.}~\bibnamefont {Mecke}}, \bibinfo {author}
  {\bibfnamefont {C.}~\bibnamefont {De~Michele}}, \bibinfo {author}
  {\bibfnamefont {T.}~\bibnamefont {Schilling}}, \bibinfo {author}
  {\bibfnamefont {M.}~\bibnamefont {Saadatfar}}, \bibinfo {author}
  {\bibfnamefont {M.}~\bibnamefont {Schr{\"o}ter}}, \bibinfo {author}
  {\bibfnamefont {G.~W.}\ \bibnamefont {Delaney}},  \emph {et~al.},\
  }\href@noop {} {\bibfield  {journal} {\bibinfo  {journal} {Europhys. Lett.}\
  }\textbf {\bibinfo {volume} {111}},\ \bibinfo {pages} {24002} (\bibinfo
  {year} {2015})}\BibitemShut {NoStop}%
\bibitem [{\citenamefont {Jiao}\ \emph {et~al.}(2010)\citenamefont {Jiao},
  \citenamefont {Stillinger},\ and\ \citenamefont {Torquato}}]{Ji10}%
  \BibitemOpen
  \bibfield  {author} {\bibinfo {author} {\bibfnamefont {Y.}~\bibnamefont
  {Jiao}}, \bibinfo {author} {\bibfnamefont {F.~H.}\ \bibnamefont
  {Stillinger}}, \ and\ \bibinfo {author} {\bibfnamefont {S.}~\bibnamefont
  {Torquato}},\ }\href@noop {} {\bibfield  {journal} {\bibinfo  {journal}
  {Phys. Rev. E}\ }\textbf {\bibinfo {volume} {81}},\ \bibinfo {pages} {041304}
  (\bibinfo {year} {2010})}\BibitemShut {NoStop}%
\bibitem [{\citenamefont {Nguyen}\ \emph {et~al.}(2014)\citenamefont {Nguyen},
  \citenamefont {Az{\'e}ma}, \citenamefont {Radjai},\ and\ \citenamefont
  {Sornay}}]{Ng14}%
  \BibitemOpen
  \bibfield  {author} {\bibinfo {author} {\bibfnamefont {D.-H.}\ \bibnamefont
  {Nguyen}}, \bibinfo {author} {\bibfnamefont {{\'E}.}~\bibnamefont
  {Az{\'e}ma}}, \bibinfo {author} {\bibfnamefont {F.}~\bibnamefont {Radjai}}, \
  and\ \bibinfo {author} {\bibfnamefont {P.}~\bibnamefont {Sornay}},\
  }\href@noop {} {\bibfield  {journal} {\bibinfo  {journal} {Phys. Rev. E}\
  }\textbf {\bibinfo {volume} {90}},\ \bibinfo {pages} {012202} (\bibinfo
  {year} {2014})}\BibitemShut {NoStop}%
\bibitem [{\citenamefont {Baule}\ \emph {et~al.}(2013)\citenamefont {Baule},
  \citenamefont {Mari}, \citenamefont {Bo}, \citenamefont {Portal},\ and\
  \citenamefont {Makse}}]{Ba13}%
  \BibitemOpen
  \bibfield  {author} {\bibinfo {author} {\bibfnamefont {A.}~\bibnamefont
  {Baule}}, \bibinfo {author} {\bibfnamefont {R.}~\bibnamefont {Mari}},
  \bibinfo {author} {\bibfnamefont {L.}~\bibnamefont {Bo}}, \bibinfo {author}
  {\bibfnamefont {L.}~\bibnamefont {Portal}}, \ and\ \bibinfo {author}
  {\bibfnamefont {H.~A.}\ \bibnamefont {Makse}},\ }\href@noop {} {\bibfield
  {journal} {\bibinfo  {journal} {Nat. Commun.}\ }\textbf {\bibinfo {volume}
  {4}},\ \bibinfo {pages} {2194} (\bibinfo {year} {2013})}\BibitemShut
  {NoStop}%
\end{thebibliography}
\end{document}